\documentclass[12pt,preprint]{aastex}
\usepackage{epsf}

\newcommand{\be}{\begin{displaymath}}
\newcommand{\ee}{\end{displaymath}}
\newcommand{\bea}{\begin{eqnarray}}
\newcommand{\eea}{\end{eqnarray}}

\shortauthors{Denissenkov et al.}
\shorttitle{Angular Momentum Transport in Solar-Type Stars}

\begin{document}

\title{ANGULAR MOMENTUM TRANSPORT IN SOLAR-TYPE STARS:
       TESTING THE TIMESCALE FOR CORE-ENVELOPE COUPLING}

\author{Pavel A. Denissenkov\altaffilmark{1,2}, Marc Pinsonneault\altaffilmark{1},
        Donald M. Terndrup\altaffilmark{1}, and Grant Newsham\altaffilmark{1}}
\altaffiltext{1}{Department of Astronomy, The Ohio State University, 4055 McPherson Laboratory,
       140 West 18th Avenue, Columbus, OH 43210, USA; dpa,\,pinsono,\,terndrup,\,newshamg@astronomy.ohio-state.edu.}
\altaffiltext{2}{Present address: Department of Physics \& Astronomy, University of Victoria,
       P.O.~Box 3055, Victoria, B.C., V8W~3P6, Canada}
 
\begin{abstract}
We critically examine the constraints on internal angular momentum transport which can be inferred from the spin down of open cluster stars.  The rotation
distribution inferred from rotation velocities and periods are consistent for larger and more recent samples, but smaller samples of rotation periods appear biased
relative to $v\sin i$ studies.  We therefore focus on whether the rotation period distributions observed in star forming regions can be evolved into the observed ones in
the Pleiades, NGC\,2516, M\,34, M\,35, M\,37, and M\,50 with plausible assumptions about star-disk coupling and 
angular momentum loss from magnetized solar-like winds.  Solid body
models are consistent with the data for low mass fully convective stars but highly inconsistent for higher mass stars where the surface convection zone can decouple
for angular momentum purposes from the radiative interior.  The Tayler-Spruit magnetic angular momentum transport mechanism, commonly employed in models of high mass
stars, predicts solid-body rotation on extremely short timescales and is therefore unlikely to operate in solar-type pre-MS and MS stars at the predicted rate.
Models with core-envelope decoupling can explain the spin down of 1.0 and 0.8 solar mass slow rotators with characteristic coupling timescales of $55\pm 25$ Myr and
$175\pm 25$ Myr respectively.  The upper envelope of the rotation distribution is more strongly coupled than the lower envelope of the rotation distribution, in accord
with theoretical predictions that the angular momentum transport timescale should be shorter for more rapidly rotating stars. Constraints imposed by the solar
rotation curve are also discussed.  We argue that neither hydrodynamic mechanisms nor our revised and less efficient prescription for the Tayler-Spruit dynamo can
reproduce both spin down and the internal solar rotation profile by themselves. It is likely that a successful model of angular momentum evolution will involve more
than one mechanism.  Further observational studies, especially of clusters younger than
100 Myr, will provide important additional constraints on the internal rotation of stars and could firmly rule out or confirm the operation of major classes of
theoretical mechanisms.
\end{abstract} 

 
\section{Introduction}
\label{sec:intro}

Rotation is an important attribute of the life of a star. When it is fast enough,
rotation can trigger various (magneto-)hydrodynamic instabilities that will drive mixing and angular 
momentum transport in the star (e.g., \citealt{z92,s99}). Unfortunately, except for the Sun, observations give
information only about surface rotation of stars. Therefore, indirect methods
have to be used to study rotation-driven transport processes in stellar interiors. 
When stars reach the main sequence (MS) their surface response to the torque from a magnetized wind depends on
the timescale for internal angular momentum transport. If open clusters are treated as an evolutionary sequence,
the assumptions of solid body (SB) and differential rotation (DR) lead to statistically distinguishable
differences in the time evolution of the distributions of cluster star rotation rates (see, for example,
\citealt{kmc95,kea97,a98}).
At late ages, models also have to be consistent with the strong coupling evident in the solar internal rotation
profile, with nearly SB rotation in the radiative core down to $\sim$0.2$\,R_\odot$ (\citealt{tst95,cea03}).

The prior estimates of the timescale for core-envelope coupling in solar-type stars usually agreed on
a short coupling time of order one Myr for the fastest rotators. However, they considerably disagreed on the coupling time
for slowly rotating solar analogs for which the estimated values varied from 10\,--\,100 Myr (e.g., \citealt{kmc95,a98}) to 0.5 Gyr 
(e.g., \citealt{iea07}). This discrepancy has been caused, on the one hand, by
not carefully treating the selection effects and statistics
and, on the other hand, by making particular assumptions about such things as the disk-locking time, initial rotation period
distribution or the relative cluster ages which have since been updated. For example, the early work
assumed that $\alpha$\,Per was 50 Myr and the Pleiades was 70 Myr old (\citealt{mb91}).
In this work we apply rigorous methods of statistical analysis to the most recent large observational datasets on 
rotation periods of low-mass stars in open clusters and extensive computational modeling of
their rotational evolution to constrain the timescale for core-envelope coupling in the slowest rotators.

In this paper we find that the timescale for core-envelope coupling depends on both rotation rate and mass,
but it is significantly longer than would be predicted if SB rotation was enforced on a short timescale.
Angular momentum transport by magnetic torques generated by the Tayler-Spruit dynamo (\citealt{s99,s02})
does not pass this key test because it always enforces SB rotation in a solar-type star, no matter how
slowly it rotates.
This disagrees with the observational evidence that
slow rotators in young clusters are most likely to possess DR (their cores rotate faster than their envelopes)
rather than follow the $P$-age relations computed using SB rotation models. Furthermore, we show that even
a revised prescription for the Tayler-Spruit dynamo proposed by \cite{dp07}, which reduces    
the effective magnetic viscosity by nearly two orders of magnitude, is still
too efficient in redistributing angular momentum in the outer parts of the radiative core;
this is inconsistent with the observed evolution of slowly rotating
solar analogues in the $P$-age plane. The revised prescription comes to a better agreement with observations when
it is supplemented by an additional angular momentum transport mechanism that should be able to operate in the inner core
on a longer timescale. Such a mechanism is also needed to assist the revised prescription in
shaping the solar SB rotation.

In our work, three different theoretical models are employed to study the rotational evolution of solar-type stars:
a simple double-zone model and two full stellar evolutionary models, one with a constant viscosity and the other
with the effective magnetic viscosities provided by the original and revised prescriptions for the Tayler-Spruit dynamo.
The models are described in detail and compared with each other in Section 2. 
Because the choice of basic model parameters is necessarily constrained by observations, we find it helpful to briefly introduce 
the samples of rotation period and $v\sin i$ data, that will be used for a more detailed analysis later in Section~3,
already at the beginning of Section~2.
In Section 3, we present
the results of our computations of rotation period distributions for solar-type stars, as they evolve from the pre-main sequence
(pre-MS) deuterium birth line to the solar age, and compare them with observations of rotation periods in
open clusters. A brief discussion and our main conclusions are given in Section 4.
 
\section{Basic Theoretical Models and Observational Data}
\label{sec:models}

For studying the angular momentum evolution of solar-type stars, it is important to specify correct initial and boundary conditions 
as well as to incorporate into the stellar evolution code three principal processes that are believed to govern changes of
their surface rotation with time: disk-locking, internal transport of angular momentum, and
angular momentum loss from the surface.
In this section, we briefly describe common ingredients of our rotating stellar evolutionary models, such as the input physics,
the initial and boundary conditions, the law used for angular momentum loss, and our considered mechanisms for internal angular momentum
transport.

Because of the complex inter-relashionship between the physical processes involved in the rotational evolution of solar-type stars
it is also important to use all available observational data that can constrain it. Fortunately, thanks to ongoing and planned ground and space
based planet transit searches (such as the Deep MMT Transit Survey, the Monitor Project, KEPLER and COROT), extensive datasets of
rotation periods for young and intermediate-age open cluster low-mass stars have been accumulating quickly during the last few years.
The data that will be used in our paper to constrain the models are summarized in Table~\ref{tab:tab1}.
Their corresponding rotation periods are plotted in Fig.~\ref{fig:f1} (crosses) for illustration.

\subsection{Common Model Ingredients}

In our full rotating evolutionary computations we employ an upgraded version of the computer code used by \cite{dv03}.
The most recent update is the adoption of Alan Irwin's improved EOS\footnote{We use the EOS code that is made publicly available 
at {\tt http://freeeos.sourceforge.net/} under the GNU General Public License.}. In addition, the energy losses due to neutrino emission are
now calculated with the code distributed by \citet{iea96}.  We use OPAL opacities (\citealt{ri92}) for
temperatures above $\sim$\,$10^4$ K, complemented by \citet{af94} data for lower temperatures.
Nuclear reaction rates are taken from the NACRE compilation (\citealt{aea99}). Gravitational settling is not included.

We accept Zahn's concept of the rotation-induced anisotropic turbulence in stellar radiative zones, with horizontal
components of the turbulent viscosity strongly dominating over those in the vertical direction (\citealt{z92}), which assumes that
the horizontal turbulence has erased the latitudinal differential rotation. This allows to consider the angular
velocity as a function of radius alone. We take into account small corrections to the stellar structure equations arising from
the distortion of equipotential surfaces by such shellular rotation (for details, see \citealt{dv03}).
We used the \cite{gn93} mixture of heavy elements.
Our code has been calibrated to reproduce the solar luminosity $L_\odot = 3.85\times 10^{33}$ erg\,s$^{-1}$
and radius $R_\odot = 6.96\times 10^{10}$ cm at the solar age of $t_\odot = 4.57$ Gyr.
This procedure yields the initial hydrogen mass fraction
$X = 0.708$ for the solar heavy-element mass fraction $Z=0.018$, and a mixing length $\alpha$ of 1.75.

\subsection{Initial and Boundary Conditions}

Our computations start on the pre-MS deuterium birth line of \cite{ps91}. For stars
with masses $M = 1\,M_\odot$ and $0.8\,M_\odot$, the estimated birth line radii are $4.8\,R_\odot$
and $2.9\,R_\odot$, respectively. 
We assume that convective regions rotate as solid bodies at all times.  Rotation at the birth line is specified
by the initial rotation period $P_0$. In practice, we begin with a non-rotating fully convective model
that is located above the birth line. This model is evolved down to the birth line where it is spun up
to $P=P_0$, which yields the initial model for our rotating evolutionary computations. 
The initial period values are taken from the observed period distributions of
solar-type stars in the youngest stellar clusters from our compiled data samples: 
the Orion Nebula cluster (hereafter, ONC or Orion),
NGC\,2264, and NGC\,2362 (Table~\ref{tab:tab1}). The Kolmogorov-Smirnov test gives high probabilities
for these distributions to have been drawn from the same real distribution (Fig.~\ref{fig:f2}).
Because stars still possess deep surface convection zones at these young ages, different ways of
achieving the same surface rotation rate at the end of this stage will have similar final outcomes.

During its approach to the zero-age main sequence (ZAMS), the protostar contracts. If its total angular momentum
$J_{\rm tot}$ conserved on the pre-MS then the star would spin up as a result of the contraction.
However, observations show that a large number of stars arriving at the ZAMS rotate slowly,
as if their angular velocity rather than angular momentum has remained constant (see Fig.~3 in \citealt{rws04}).
The most plausible explanation is that the interaction of protostars with their accretion disks extracts
angular momentum from the central object, reducing or preventing spin-up as they contract.
This can occur from magnetic coupling between the protostar and disk (\citealt{k91,shea94}) or through
an enhanced wind (\citealt{mp05}).
As in most other works (e.g., \citealt{kea97,bfa97,a98,tpt02}), we model the interaction between the protostar and
accretion disk in a simple way, assuming that during an initial time interval
$0\leq t\leq\tau_{\rm d}$ the interaction maintains $P(t) = P_0$. 
The disk-locking time $\tau_{\rm d}$ is considered as a free parameter.

In our stellar evolution code the transport of angular momentum is treated as a diffusion process
described by
\bea
\frac{d}{dt}(r^2\Omega) = \frac{\partial}{\partial M_r}\left[(4\pi r^2\rho)^2r^2\nu\frac{\partial\Omega}{\partial M_r}\right],
\label{eq:amt}
\eea
where $d/dt$ is a derivative taken at a constant $M_r$, and 
$\nu$ is a viscosity whose physical nature has yet to be identified in real stars.
Equation (\ref{eq:amt}) needs two boundary conditions. A natural initial condition for it is that the fully convective
birth line model rotates as a solid body with $\Omega(0,M_r)=\Omega_{\rm e}(0)=2\pi/P_0$, where $\Omega_{\rm e} = 2\pi/P$
is the angular velocity of star's convective envelope. The inner SB-rotation boundary condition
$\partial \Omega/\partial M_r = 0$
is applied at the surface of the innermost mass shell used in our computations, at $M_r\approx 0$.
To derive the outer boundary condition that has to be applied at the bottom of convective envelope $M_r=M_{\rm bce}$, we integrate
equation (\ref{eq:amt}) from $M_r=0$ to $M_r=M_{\rm bce}$, taking into account that the time derivative of the angular momentum of radiative core is
$$  
\dot{J}_{\rm c}=\frac{d}{dt}\ \frac{2}{3}\int_0^{M_{\rm bce}}r^2\Omega\,dM_r =
\frac{2}{3}\,r_{\rm bce}^2\Omega_{\rm e}\dot{M}_{\rm bce}+\frac{2}{3}\int_0^{M_{\rm bce}}\frac{d}{dt}(r^2\Omega)\,dM_r =
\dot{J}_{\rm tot}-\dot{J}_{\rm e}.
$$  
Here, $J_{\rm e} = I_{\rm e}\Omega_{\rm e}$ is the angular momentum of convective envelope, $I_{\rm e}$ being the envelope's moment of inertia.
Finally, we obtain the following upper boundary condition:
\bea
I_{\rm e}\dot{\Omega}_{\rm e} = \dot{J}_{\rm tot} - \dot{I}_{\rm e}\Omega_{\rm e} -
\frac{2}{3}\,r_{\rm bce}^2\Omega_{\rm e}\dot{M}_{\rm bce} -
\frac{2}{3}\left[(4\pi r^2\rho)^2r^2\nu\frac{\partial\Omega}{\partial M_r}\right]_{M_{\rm bce}},
\label{eq:upperbc}
\eea
in which $\dot{J}_{\rm tot}$ is the rate of angular momentum loss from the stellar surface.

\subsection{Angular Momentum Loss and Internal Transport}
\label{sec:amloss}

For the angular momentum loss from the surface, we adopt the magnetized stellar wind prescription of \cite{kea97}:
\bea
\dot{J}_{\rm tot} = -K_{\rm w}\sqrt{\frac{R/R_\odot}{M/M_\odot}}
\,\min\left(\Omega_{\rm e}\Omega_{\rm sat}^2,\,\Omega_{\rm e}^3\right).
\label{eq:jdot}
\eea
Here, $\Omega_{\rm sat}$ is the velocity at which the wind is saturated.
The parameter $\Omega_{\rm sat}$ is known to strongly depend on the stellar mass (e.g., \citealt{aps03}).
In Fig.~\ref{fig:f3}, we have used upper 90th percentiles of the $\Omega_{\rm e}$ distributions
for our cluster sample to adjust the appropriate values of $\Omega_{\rm sat}$ for the three mass bins that
we use in this study. For internal angular momentum transport we assume that fast rotators behave as SB rotators,
a result both of theoretical calculations (\citealt{kea97,spt00}) and indicated by prior spin down
studies (\citealt{iea07}; there were prior claims by the Monitor group).
The error bars to the percentiles were calculated using bootstrap simulations
by generating 1000 sampling distributions for each cluster. Our adjusted $\Omega_{\rm sat}$ values (shown in each panel in Fig.~\ref{fig:f3})
are close to those reported by \cite{aps03}. In particular, we have found that the same value of
$\Omega_{\rm sat} = 8\,\Omega_\odot$ can be used for either of our ``solar-type'' mass bins, $0.7\leq M/M_\odot < 0.9$ and
$0.9\leq M/M_\odot\leq 1.1$, to reproduce reasonably well their corresponding 90th percentiles.
This value lies between the estimates $\Omega_{\rm sat} = 6.4\,\Omega_\odot$ and $\Omega_{\rm sat} = 10.5\,\Omega_\odot$
obtained by \cite{aps03} for stars with $M = 0.8\,M_\odot$ and $M = 1.0\,M_\odot$, respectively.
For the ``fully convective'' $M\leq 0.4\,M_\odot$ bin, we have adjusted the parameter $\Omega_{\rm sat} = 2.5\,\Omega_\odot$.
It can be compared with the value of $1.8\,\Omega_\odot$ used by \cite{aps03}
as a saturation threshold for their $0.4\,M_\odot$ model star. Note that in all our computations we use
stellar models with masses appropriate for the considered mass bins, namely $0.3\,M_\odot$, $0.8\,M_\odot$, and $1.0\,M_\odot$.
The parameter $K_{\rm w}$ in equation (\ref{eq:jdot}) is calibrated by requiring that our $0.8\,M_\odot$ and $1.0\,M_\odot$ models have 
$\Omega_{\rm e}=\Omega_\odot = 2.86\times 10^{-6}$\,rad\,s$^{-1}$ ($P_\odot = 25.4$ days)
at the solar age. For stars with $M\la 0.4\,M_\odot$, we use our $0.3\,M_\odot$ model and the solar calibrated value of 
$K_{\rm w}\approx 3.19\times 10^{47}$\,cm$^2$\,g\,s$^{-2}$ that has produced the solid curve in Fig.~\ref{fig:f3}a.

\subsubsection{Constant Viscosity}

In this section, we consider angular momentum transport in rotating stellar evolutionary models with a constant viscosity $\nu(t,M_r)=\nu_0$.
Although the assumption of constant viscosity does not reveal the physical mechanism
responsible for the transport of angular momentum in solar-type stars, it nevertheless permits   
us to estimate both an instructive minimum value $(\nu_0)_{\rm min}$ that still results in
the solar SB rotation and a maximum value $(\nu_0)_{\rm max}$ such that   
the rotational evolution with any value in excess of it looks identical to that with $\nu_0 = (\nu_0)_{\rm max}$. 
We have computed the evolution of a rotating
$1\,M_\odot$ star for two combinations of the disk-locking time (in Myr) and initial
rotation period (in days): $(\tau_{\rm d},P_0) = (6,8)$ and $(2,3)$. For either of these
combinations, the computations have been done for the same set of values of $\Omega_{\rm sat} = 8\,\Omega_\odot$ and
$\nu_0 = 2.5\times 10^4,\,5\times 10^4,\,10^5,\,2.5\times 10^5,\,\,5\times 10^5,\,10^6,\,10^7$, and $10^8$
cm$^2$\,s$^{-1}$. Results are presented in Fig.~\ref{fig:f4} with solid and dashed curves for the first and
second combination of initial conditions, respectively.
Models with $\nu_0\leq (\nu_0)_{\rm min}\approx 5\times 10^4$ cm$^2$\,s$^{-1}$ have
a residual DR at the solar age inconsistent with helioseismic data. On the other hand,
the rotational evolution of models with $\nu_0\geq (\nu_0)_{\rm max}\approx 10^6$ cm$^2$\,s$^{-1}$
is almost indistinguishable from one another.    

\subsubsection{Effective Magnetic Viscosities from the Tayler-Spruit Dynamo}

\cite{s99,s02} has elaborated upon the finding by \cite{t73} that
toroidal magnetic fields frozen into plasma in a stellar radiative zone are always subject to 
a pinch-type instability. A release of magnetic energy by this instability 
causes concentric magnetic rings to slide sideways, mainly horizontally,
along the equipotential surfaces and, to some extent, also along the radius.
Magnetic induction makes it possible for the unstable radial displacement
to produce a weak poloidal field $B_r$ at the expense of toroidal magnetic field $B_\varphi$.
If the radiative zone rotates differentially, the poloidal field can be stretched
around the rotation axis into a new toroidal field that will again be subject to
the Tayler instability. Spruit's original idea was that
these consecutive poloidal/toroidal field generations might sustain each other under
certain circumstances, thus forming a dynamo loop.
The Tayler-Spruit dynamo could drive some mixing through unstable radial displacements
with an effective diffusion coefficient $\eta_{\rm e}$, as well as some angular momentum
transport by magnetic torques proportional to the product $B_rB_\varphi$
with an effective viscosity $\nu_{\rm e}$. However, when considering circumstances under
which his proposed dynamo might work, Spruit did not take into account a reduction of
the unstable long-scale horizontal displacement by the Coriolis force, therefore overestimating
$\eta_{\rm e}$ and $\nu_{\rm e}$. The Coriolis force has properly been taken into account
and the transport coefficients have been revised correspondingly by \cite{dp07}.
Further model developments related to the Tayler-Spruit dynamo were proposed by
\cite{zbm07} and \cite{rgs09}.

Whereas the original prescription for the Tayler-Spruit dynamo has been shown
to produce a solar rotation profile in agreement with helioseismic data
(\citealt{eea05}), the revised prescription fails to do so alone (\citealt{dp07}).            
One of the objectives of the present work is to see if these prescriptions      
can yield the evolution of surface rotation consistent with the rotation period data
for solar-type stars in open clusters.

Let us summarize the basic equations for the magnetic transport coefficients in the Spruit
mechanism. For details, the reader is referred to the paper of \cite{dp07}.
For Spruit's original prescription, it is convenient to present the effective
magnetic diffusivity and viscosity in the following forms:
\bea
\eta_{\rm e} = \alpha\,Ky^3,\ \ \mbox{and}\ \
\nu_{\rm e} = \alpha\,K\frac{N_T^2+N_\mu^2}{\Omega^2q^2}y^2.
\label{eq:mm04}
\eea
Here, $q=|\partial\ln\Omega/\partial\ln r|$ is the rotational shear, and $y$ is a solution of the 4th order algebraic equation
\bea
\alpha\,y^4 - \alpha\,y^3 + \beta\,y - 2 = 0,
\label{eq:4th}
\eea
where
\bea
\alpha = r^2\frac{\Omega^7\,q^4}{K(N_T^2+N_\mu^2)^3},\ \ \mbox{and}\ \ \beta = 2\frac{N_\mu^2}{N_T^2+N_\mu^2}
\label{eq:4thcoeff}
\eea
are dimensionless coefficients. In equations (\ref{eq:mm04}) and (\ref{eq:4thcoeff}), we have used
standard notations for the thermal diffusivity
\bea
K = \frac{4acT^3}{3\kappa\rho^2C_P},
\eea
where $\kappa$ and $C_P$ represent the opacity and the specific heat at constant pressure, respectively, and
for the $T$- and $\mu$-component of the square of the  Brunt-V\"{a}is\"{a}la (buoyancy) frequency
\be
N_T^2 = \frac{g\delta}{H_P}(\nabla_{{\rm ad}}-\nabla_{{\rm rad}}),
\ \ \mbox{and}\ \
N_\mu^2 = g\varphi\,\left|\frac{\partial\ln\mu}{\partial r}\right|.
\ee
In the last expressions, $g$ is the local gravity, $H_P$ is the pressure scale height,
$\nabla_{\rm ad}$ and $\nabla_{\rm rad}$ are the adiabatic and radiative temperature gradients
(logarithmic and with respect to pressure), and
$\mu$ is the mean molecular weight.
The quantities
$\delta =
-\left(\partial\ln\rho/\partial\ln T\right)_{P,\mu}$ and
$\varphi = \left(\partial\ln\rho/\partial\ln\mu\right)_{P,T}$
are determined by the equation of state. In particular, for the perfect gas law $\delta = \varphi = 1$.

When using the effective magnetic viscosity $\nu = \nu_{\rm e}$ in the angular momentum transport equation
(\ref{eq:amt}), it is important to remember that the Tayler-Spruit dynamo
keeps operating only as long as the effective diffusivity $\eta_{\rm e}$ remains
larger than the magnetic diffusivity $\eta_{\rm mag}$. When $\eta_{\rm e}$ approaches $\eta_{\rm mag}$
the poloidal magnetic field decays through Ohmic dissipation faster than it is
generated by the unstable radial displacements of magnetic rings. Although theory says nothing about the behavior of
$\nu_{\rm e}$ in this limit, we can formally estimate a minimum value of the effective magnetic viscosity 
that is reached when $\eta_{\rm e}=\eta_{\rm mag}$.
To do this, we replace $\eta_{\rm e}$ with $\eta_{\rm mag}$ in the first of equations (\ref{eq:mm04}) and then solve
equations (\ref{eq:mm04}\,--\,\ref{eq:4thcoeff}) with respect to $q$ and $\nu_{\rm e}$.
The effective magnetic viscosity approaches its minimum value of
\bea
(\nu_{\rm e})_{\rm min} = 6.96\times 10^7\left(\frac{r}{R_\odot}\right)
\left(\frac{2+\varepsilon}{\beta +\varepsilon}\right)^{1/2}\left(\frac{\eta_{\rm mag}}{10^3}\right)^{1/2}
\left(\frac{N_T^2+N_\mu^2}{10^{-6}}\right)^{-1/2}
\left(\frac{\Omega}{10^{-5}}\right)^{3/2}\ \mbox{cm}^2\,\mbox{s}^{-1}
\label{eq:nuemin}
\eea
when the shear is reduced to
\bea
q_{\rm min} = 0.379\,\left(\frac{r}{R_\odot}\right)^{-1/2}
\left(\frac{\beta+\varepsilon}{2 +\varepsilon}\right)^{3/4}\left(\frac{\eta_{\rm mag}}{10^3}\right)^{1/4}
\left(\frac{N_T^2+N_\mu^2}{10^{-6}}\right)^{3/4}
\left(\frac{\Omega}{10^{-5}}\right)^{-7/4}.
\label{eq:qmin}
\eea
The normalizations in the last two equations, for which we have used values typical for
the solar interior, assume that all quantities are expressed in cgs units.
We have also introduced the reciprocal of the Roberts number $\varepsilon = \eta_{\rm mag}/K\ll 1$.

The ratio $\beta$ defined by the second of equations (\ref{eq:4thcoeff}) remains small
$(\beta\ll 1)$ everywhere in the star until the age of $\sim$\,30 Myr because nuclear reactions
have not yet built up a sufficiently strong $\mu$-gradient in the stellar core. Given that by this
age $\Omega_{\rm e}$ approaches its maximum value
that turns out to exceed $\sim$\,$3\Omega_\odot\approx 10^{-5}$ rad\,s$^{-1}$ in most interesting cases
(see Section \ref{sec:confrwithobs}, and panel a in Fig.~\ref{fig:f3}), 
it is obvious from equation (\ref{eq:nuemin}) that
$(\nu_{\rm e})_{\rm min}$ will be of order $10^8$\,cm$^2$\,s$^{-1}$ everywhere in our model star during the first
tens Myr of its evolution. Furthermore, since $\beta$ continues to remain very small outside 
the radius $r\sim$\,$0.2\,R_\odot$, where $N_\mu^2\ll N_T^2$,
up to the solar age, we can expect that $(\nu_{\rm e})_{\rm min}$
will keep values of order $10^6$\,--\,$10^8$ in the outer part of the radiative core even at older ages.
These expectations are confirmed by our detailed computations. Taking into account that     
equation (\ref{eq:nuemin}) gives only a lower limit for the effective magnetic viscosity and
the inner core with $r\la 0.2\,R_\odot$ contributes less than 10\% to the total moment of inertia of
the radiative core, we anticipate that Spruit's original prescription should always bring about
the $P$-age relations characteristic of SB rotators, 
like those obtained with our constant viscosity model for $\nu_0 > (\nu_0)_{\rm max}$.
However, we have to ensure that in our computations the viscosity declines abruptly as soon as
the shear $q$ is reduced below its
critical value given by equation (\ref{eq:qmin}). This requirement may leave some residual DR
in the core of our final solar model.

In our full rotating stellar evolutionary models we have used an approximate computational method.                       
It assumes that, as soon as a radiative core develops in a pre-MS star and its rotation profile begins to deviate
from uniform one, the effective magnetic viscosity is large enough everywhere
in the core to potentially restore its uniform rotation. Our test computations confirm this.
However, when the redistribution of angular momentum by
magnetic torques has led to $q\approx q_{\rm min}$, the Tayler-Spruit dynamo ceases to work, and
its related viscosity should be replaced with the
molecular one $\nu_{\rm mol}\ll (\nu_{\rm e})_{\rm min}$. As we have noted, theory does not describe
how this transition from $(\nu_{\rm e})_{\rm min}$ to $\nu_{\rm mol}$ occurs. 
When an increase of $\Omega_{\rm e}$ at $t > \tau{\rm d}$ caused by the residual pre-MS contraction of the star
and its angular momentum conservation finally gives way to an $\Omega_{\rm e}$ decrease due to the surface loss of
angular momentum with the magnetized stellar wind, a much stronger DR tends to accumulate in the core.
However, as soon as the shear exceeds its critical value $q_{\rm min}$, a large viscosity of order
$(\nu_{\rm e})_{\rm min}$ will resume the redistribution of angular momentum by magnetic torques smoothing out the
angular velocity gradient until $q$ drops below $q_{\rm min}$ again. Following this qualitative picture,
we put into equation (\ref{eq:amt}) $\nu = (\nu_{\rm e})_{\rm min}$ multiplied by
an exponential factor that cancels $\nu$ when $q$ approaches $q_{\rm min}$.

Fig.~\ref{fig:f5} compares the results of our computations using this method
for $(\tau_{\rm d},\,P_0) = (6,\,8),\ (6,\,30),\ \mbox{and}\ (2,\,3)$ (solid curves) with results that we obtained
for the same initial conditions but applying the constant viscosities $\nu_0 = 10^6$ and $10^7$ cm$^2$\,s$^{-1}$ (dashed curves).
We have $\Omega_{\rm sat} = 8\,\Omega_\odot$ throughout.                 
The comparison shows that, for Spruit's original prescription, the rotation period evolution of the Sun coincides
with that of $\nu_0 = 10^7$\,cm$^2$\,s$^{-1} > (\nu_0)_{\rm max}$; i.e., it always corresponds to the SB rotation case.
However, the solid curve in the bottom panel in Fig.~\ref{fig:f6} demonstrates that, 
unlike the constant viscosity model with $\nu_0 > (\nu_0)_{\rm max}$, 
our solar model computed using Spruit's original prescription does contain a small differentially rotating
core, as we expected. Unfortunately, its presence cannot be revealed by available helioseismic data
(see, however, \citealt{gea07}). This result agrees with that reported by \cite{eea05}.

For Spruit's revised prescription (\citealt{dp07}), the effective magnetic diffusivity and viscosity are given by the following equations:
\bea
\eta_{\rm e} = 2K\frac{\Omega^2 q^2 - N_\mu^2}{N_T^2 + N_\mu^2 - \Omega^2 q^2},\ \ \mbox{and}\ \  
\nu_{\rm e} = \left(\frac{r^2\Omega\eta_{\rm e}^2}{q^2}\right)^{1/3}.
\label{eq:rev}
\eea
After the substitution of $\eta_{\rm e}=\eta_{\rm mag}$ into the first of these equations, we find that the revised
Tayler-Spruit dynamo ceases to work when the shear approaches a critical value
\bea
q_{\rm min} = 10^2\,\left(\frac{\varepsilon}{2+\varepsilon}\,\frac{N_T^2}{10^{-6}}+\frac{N_\mu^2}{10^{-6}}\right)^{1/2}
\left(\frac{\Omega}{10^{-5}}\right)^{-1}.
\label{eq:qminrev}
\eea
A comparison of coefficients in equations (\ref{eq:qmin}) and (\ref{eq:qminrev}) shows
that in the second case the residual DR in the solar model is expected to be much stronger.
A minimum value of $\nu_{\rm e}$ can be estimated from the second of equations (\ref{eq:rev})
in which we put $\eta_{\rm e} = \eta_{\rm mag}$ and $q = q_{\rm min}$. As a result, we get
\bea
(\nu_{\rm e})_{\rm min} = 1.69\times 10^6 \left(\frac{r}{R_\odot}\right)^{2/3}
\left(\frac{\eta_{\rm mag}}{10^3}\right)^{2/3}
\left(\frac{\varepsilon}{2+\varepsilon}\,\frac{N_T^2}{10^{-6}}+\frac{N_\mu^2}{10^{-6}}\right)^{-1/3}
\left(\frac{\Omega}{10^{-5}}\right)\ \mbox{cm}^2\,\mbox{s}^{-1},
\label{eq:nueminrev}
\eea
where all numbers are again expressed in cgs units.

It is instructive to compare the minimum values of the effective magnetic viscosity
$(\nu_{\rm e})_{\rm min}^{(\ref{eq:nuemin})}$ and $(\nu_{\rm e})_{\rm min}^{(\ref{eq:nueminrev})}$ given by
equations (\ref{eq:nuemin}) and (\ref{eq:nueminrev}), respectively. Their ratio is
\bea
\frac{(\nu_{\rm e})_{\rm min}^{(\ref{eq:nuemin})}}{(\nu_{\rm e})_{\rm min}^{(\ref{eq:nueminrev})}} =
41.2\,\left(\frac{r}{R_\odot}\right)^{1/3}\left(\frac{2+\varepsilon}{\varepsilon}\right)^{1/6}
\left(\frac{\eta_{\rm mag}}{10^3}\right)^{-1/6}
\left(\frac{N_T^2}{10^{-6}}\right)^{-1/6}\left(\frac{\Omega}{10^{-5}}\right)^{1/2}.
\label{eq:nueratio}
\eea
The last two equations show that, even though the revised value of
$(\nu_{\rm e})_{\rm min}$ is less than the original one, it is still large enough for us to employ the same
computational method that we used to study the evolution of rotating solar-type stars with Spruit's original prescription.

In Fig.~\ref{fig:f6}, results of our application of Spruit's original and revised prescriptions in full
evolutionary computations of rotating solar models are compared with each other as well as with
a rotating model with no internal transport of angular momentum (solid, dashed, and dotted
curve, respectively). We confirm the conclusion made by \cite{dp07}, who used a crude approximation
$$
\Omega(r) = \Omega_\odot + \int_r^{r_{\rm bce}}\Omega\, q_{\rm min}\,\frac{dr}{r}
$$
to construct an $\Omega$-profile in their model of the present-day Sun, that the revised prescription for the Tayler-Spruit
dynamo produces a solar model with a large rapidly rotating core, in contradiction with helioseismic data  
(compare the dashed curve with the filled circles representing observational data from \citealt{cea03} in bottom panel).
Our new result obtained here is that, in spite of this, the revised prescription leads to the rotational evolution
that has even shorter periods at ages older than $\sim$\,30 Myr than the SB rotational evolution obtained
with Spruit's original prescription (compare the dashed and solid curves in top panel).
This behavior is explained as follows. As we anticipated, the values of $(\nu_{\rm e})_{\rm min}$ given by
equation (\ref{eq:nueminrev}) indeed turned out to be large enough for the transport of angular momentum from the core
to envelope to occur on a short ``SB rotation'' timescale like those we got in our constant viscosity
models with $\nu_0 > (\nu_0)_{\rm max}\approx 10^6$ cm$^2$\,s$^{-1}$. However, because of an early build-up of
a steep critical $\Omega$-profile (like that shown with dashed curve in bottom panel), below which
the Tayler-Spruit dynamo ceases to work, the amount of the core's angular momentum available for
transport to the envelope is diminished more and more as the evolution proceeds beyond an age of $\sim$\,30 Myr.
As a result, if we used the value of the stellar wind parameter $K_{\rm w}=2.8\times 10^{47}$\,cm$^2$\,g\,s calibrated for
Spruit's original prescription then the $P$-age relation for the revised prescription would be nearly parallel
to the solid curve in top panel but it would be located at longer periods, hence the solar rotation period
would be overestimated. To match the solar surface rotation, we had to reduce $K_{\rm w}$ to $1.25\times 10^{47}$\,cm$^2$\,g\,s for
the revised prescription. This has shifted the $P$-age relation toward its location shown with the dashed curve.
Note that we have also done test computations in which $\eta_{\rm e}$ from the first of equations
(\ref{eq:rev}) was used instead of $\eta_{\rm mag}$ to estimate $\nu_{\rm e}$. We have not found
noticeable differences with the results obtained using the approximate method.

\subsection{Double-Zone Model}

The two zone model was originally proposed by \cite{mcg91} and it has since been employed by many others
(e.g., \citealt{mb91,kmc95,sl97,a98,iea07}). 
Angular momentum transport between the radiative core and convective envelope is parameterized as follows.
The core and envelope, with moments of inertia and rotation rates $I_{\rm c}$, $\Omega_{\rm c}$, $I_{\rm e}$
and $\Omega_{\rm e}$ respectively, are assumed to rotate as solid bodies.
If $\Omega_{\rm c} > \Omega_{\rm e}$ then the maximum angular momentum that can be transfered from the core
to the envelope $(\Delta J)_{\rm max}$ is estimated as a difference between the core's initial angular momentum 
$J_{\rm c} = I_{\rm c}\Omega_{\rm c}$
and the angular momentum $J_{\rm c,eq} = I_{\rm c}\Omega_{\rm eq}$ 
the core will have when its angular velocity becomes equal to that of the envelope.
This final equilibrium velocity $\Omega_{\rm eq}$ is determined from
the angular momentum conservation: $J_{\rm c} + J_{\rm e} = J_{\rm c,eq} + J_{\rm e,eq}$, or
$I_{\rm c}\Omega_{\rm c} + I_{\rm e}\Omega_{\rm e} = (I_{\rm c} + I_{\rm e})\Omega_{\rm eq}$. So, we have
$$
(\Delta J)_{\rm max} = \frac{I_{\rm c}\,I_{\rm e}}{I_{\rm c}+I_{\rm e}}(\Omega_{\rm c}-\Omega_{\rm e}).
$$
As a free parameter, the double-zone model uses the core/envelope coupling time $\tau_{\rm c}$ that defines
the rate $(\Delta J)_{\rm max}/\tau_{\rm c}$ with which the angular momentum is transfered from the core to
the envelope. An advantage of this model is that one needs to know only how $I_{\rm c}$, $I_{\rm e}$,
$r_{\rm bce}$, and $M_{\rm bce}$ are changing with time for a particular star
(the last two quantities are required to take into account a displacement of
the bottom of convective envelope during the angular momentum transfer).
If this information is available (from full evolutionary computations) then the star's rotational evolution
is obtained as a solution of the following system of ODEs:
\bea
\frac{dJ_{\rm c}}{dt} & = & -\frac{(\Delta J)_{\rm max}}{\tau_{\rm c}} +
\frac{2}{3}\,r_{\rm bce}^2\Omega_{\rm e}\dot{M}_{\rm bce}, \\
\frac{dJ_{\rm e}}{dt} & = & \frac{(\Delta J)_{\rm max}}{\tau_{\rm c}} -
\frac{2}{3}\,r_{\rm bce}^2\Omega_{\rm e}\dot{M}_{\rm bce} + \dot{J}_{\rm tot},
\label{eq:2zone}
\eea
where $\dot{J}_{\rm tot}$ should be replaced by expression (\ref{eq:jdot}).

Because of its simplicity the double-zone model can be used for Monte Carlo simulations and other statistical studies,
like those conducted in next section. However, before doing this we want to relate
the constant viscosity model and the double-zone model to one another through their parameters $\nu_0$ and $\tau_{\rm c}$.
In Fig.~\ref{fig:f7}, values of the coupling time are adjusted so that
the double-zone model simulates some of the $P$-age relations that we computed
with the constant viscosity model using the same initial conditions.
We have established the following approximate\footnote{As the morphology of $P$-age relations
is slightly different for these models, the adjusted values of $\tau_{\rm c}$ are uncertain
within $\sim$\,10\%.} correspondence between $\nu_0$ (in cm$^2$\,s$^{-1}$)
and $\tau_{\rm c}$ (in Myr): $(\nu_0,\,\tau_{\rm c}) = (2.5\times 10^4,\,90)$, $(5\times 10^4,\,40)$,
$(10^5,\,20)$, $(2.5\times 10^5,\,7)$, and $(10^6,\,1)$. This means that the double-zone models with the coupling time longer than
$\sim$\,40 Myr correspond to the constant viscosity models that, by the solar age, still possess residual 
DR inconsistent with helioseismic data. On the other hand, the double-zone models with
$\tau_{\rm c}\la 1$ Myr reproduce the $P$-age relations for SB rotators (cf. \citealt{a98}); they are equivalent
to the constant viscosity models with $\nu_0 > (\nu_0)_{\rm max}\approx 10^6$ cm$^2$\,s$^{-1}$ as well as
to the models in which angular momentum is redistributed by magnetic torques generated by the Tayler-Spruit dynamo
in its original prescription.

\section{Comparison with Observations}
\label{sec:confrwithobs}

From a theoretical standpoint, the rotational evolution of low-mass stars is a very complex process.
It depends on a number of parameters, such as the initial rotation period $P_0$, disk-locking time
$\tau_{\rm d}$, angular velocity threshold for the magnetized wind saturation $\Omega_{\rm sat}$, 
and the rate of angular momentum redistribution    
expressed in terms of the core/envelope coupling time $\tau_{\rm c}$, constant viscosity
$\nu_0$, or as a function of stellar structure and other parameters when a particular physical mechanism is chosen to describe it.
Because of this complexity, the progress in this field is primarily driven by constantly accumulating and improving observational
data and their statistical analyses. 

A number of physical mechanisms have been proposed to explain
the solar SB rotation, such as the smoothing of DR by the back reaction of the Lorentz force
emerging from the generation of a toroidal magnetic field by shearing of a preexisting poloidal field
(\citealt{mw87,chmg93}), angular momentum redistribution by magnetic torques generated by the Tayler-Spruit 
dynamo (\citealt{eea05}), or by internal gravity waves excited by turbulent eddies in the solar convective envelope
(\citealt{cht05}; see, however, \citealt{dea08}). However, none of these mechanisms has been shown to agree or disagree with
available rotation period data for solar-type stars in open clusters. In this paper, we subject the Tayler-Spruit
dynamo to such an observational test. The other mechanisms will be tested in our forthcoming papers.
We will start with Spruit's original prescription. As it always results in the $P$-age relations similar
to those obtained with the double-zone model having a short coupling time of order 1 Myr (compare
Figs.~\ref{fig:f5} and \ref{fig:f7}), we will use the latter as its substitute model. The main advantage of
this replacement is that the double-zone model computations are very fast, therefore they can effectively be used
to perform extensive parameter-space investigations.

\begin{figure}
\epsfxsize=13cm
\epsffile [60 180 480 695] {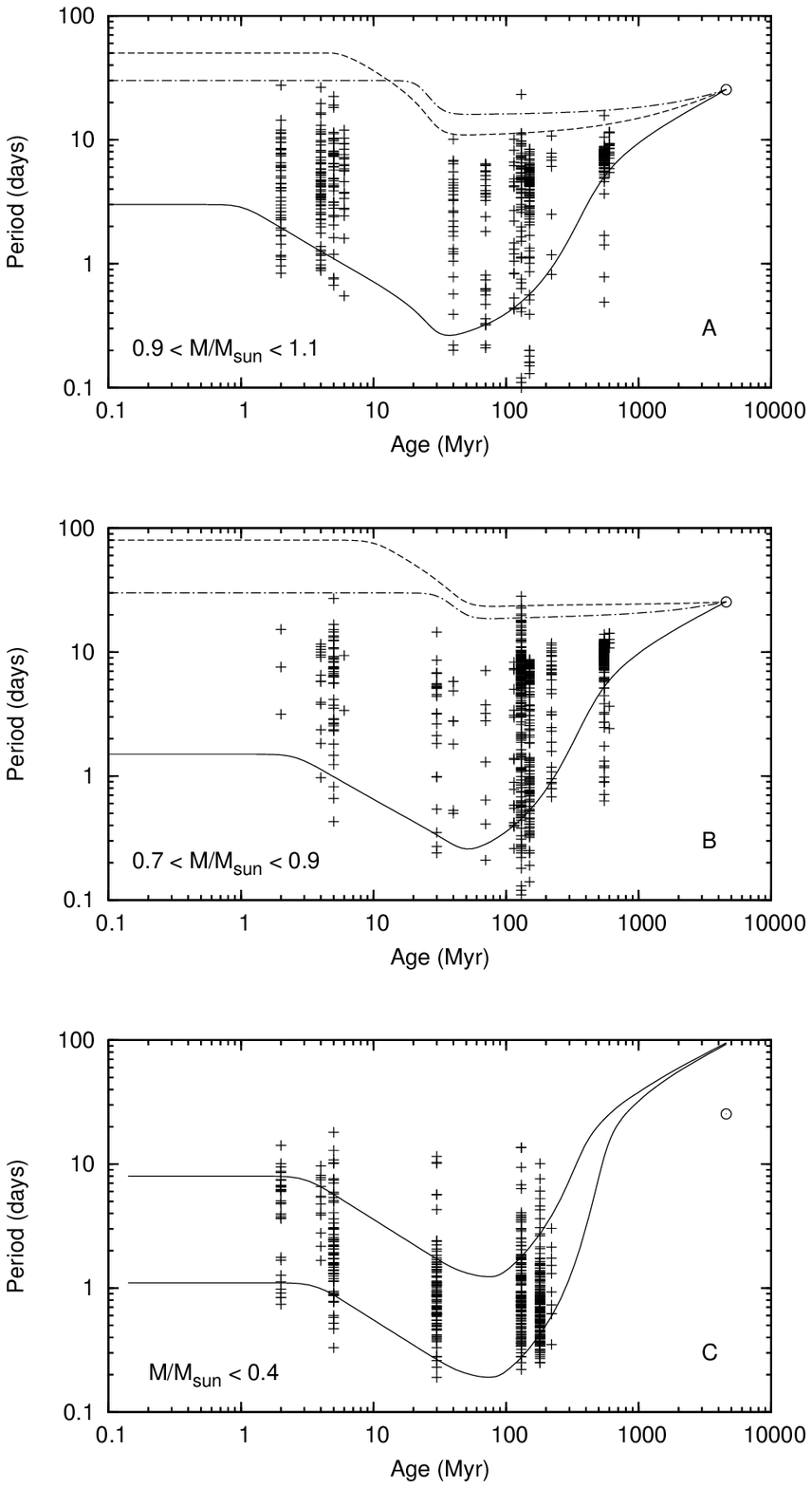}
\caption{Rotation period data (crosses) for the three mass bins used in this paper (see Table~\ref{tab:tab1}).
         Curves represent the SB rotational evolution computed with different values of  
         the initial period and disk-locking time using the double-zone model and common model ingredients
         that are described and discussed in text later.
         }
\label{fig:f1}
\end{figure}

Our first step in reducing the number of free model parameters is the adjustment of the value of $\Omega_{\rm sat}$ (see Section \ref{sec:amloss}).
For the SB rotating ($\tau_{\rm c} = 1$ Myr) double-zone model to reproduce as close as possible the upper 90th percentiles simultaneously for all of
our compiled $\Omega_{\rm e}$ data for each of the three mass bins, we had to choose $\Omega_{\rm sat} = 8\,\Omega_\odot$
for our two ``solar mass'' bins and $\Omega_{\rm sat} = 2.5\,\Omega_\odot$ for the ``fully convective'' mass bin 
(upper solid curves in Fig.~\ref{fig:f3}, these curves are also plotted in Fig.~\ref{fig:f1}, with $\Omega_{\rm e}$ being
transformed back to $P$). This procedure assumes quite naturally that the most rapidly rotating stars in the samples evolve
as SB rotators. In fact, it is impossible to construct a double-zone model with DR that would fit 
the 90th percentiles for the intermediate-age ($\sim$100 Myr old) clusters,
those located immediately behind the maxima on the $\Omega_{\rm e}$ vs. age curves, without making unreasonable
assumptions about its parameters.

It turns out that the lower 10th percentiles for the stars with $M\leq 0.4\,M_\odot$ can also be approximated reasonably well
with a SB rotation evolution curve (the lower and upper solid curves in Fig.~\ref{fig:f3}c and Fig.~\ref{fig:f1}c,
respectively). This result is expected because a turbulent eddy viscosity in these fully convective stars
should redistribute angular momentum very quickly (e.g., see \citealt{tpt02}). 
Having said that, we should mention a recent evidence that casts some doubt
on this simple interpretation. The first period measurements for five stars with masses below
$0.5\,M_\odot$ in Praesepe (\citealt{se07}) appear to disagree with the predicted model trends,
but this sample is not of sufficient size to draw firm conclusions.

On the contrary, it turns out to be very difficult for a SB rotation evolution curve to approach all the lower 10th percentiles 
for stars in the mass bins centered at $1.0\,M_\odot$ and $0.8\,M_\odot$ (dashed curves in panels a and b in Fig.~\ref{fig:f3}),
unless one takes a disk-locking time well in excess of 20 Myr or starts with a very slowly rotating star.
If the double-zone model parameters were not constrained by observational data but could be chosen at our will then any mechanism of angular momentum
transport that persistently produces and maintains SB rotation, in particular Spruit's original prescription,
could easily be brought in agreement with the open cluster rotation period data for slowly rotating solar-type stars.
Indeed, in this case one could simply choose appropriate combinations of the initial period
and disk-locking time, one or both of which having to be sufficiently long, such that the double-zone model
with SB rotation ($\tau_{\rm c} = 1$ Myr) and with those parameters applied would embrace all the periods for slow rotators,
no matter how long they are (dashed and dot-dashed curves in panels a and b in Fig.~\ref{fig:f1}).

However, observations do not allow such an arbitrary choice of parameters of a rotational evolution model for solar-type stars.
In particular, the typical disk-locking time has been estimated to lie in a range between 2 and 10 Myr.
This result is based on the measuring of such diagnostics of the presence of a circumstellar disk
around a pre-MS star as an IR excess, that traces dust, or an H$_\alpha$ emission line width that traces accretion
(\citealt{h05,ll05,bea06,jea06,dea07}). A recent statistical analysis of available data on
the pre-MS circumstellar disks has led \cite{m09} to the conclusion that
``the fraction of young stars with optically thick primordial disks and/or those which
show spectroscopic evidence for accretion appears to approximately follow an exponential decay
with characteristic time $\sim$2.5 Myr''. Taken at its face value, this means that, on average, only 10\% of
active disks around pre-MS solar-type stars are expected to survive by the age of 5.8 Myr. However, the dispersion of the disk life times
can be quite large. For instance, \cite{saea09} have found that $\sim$50\% of 18 members of the $\eta$ Cham cluster
studied by them still show signatures of a circumstellar disk by the age of 8 Myr.
Therefore, in our models we will use the maximum disk-locking time $(\tau_{\rm d})_{\rm max} = 20$ Myr as
a safe upper limit.

With the maximum disk-locking time limited by the value of 10 Myr or even 20 Myr,
the SB rotational evolution can explain the longest periods in the intermediate-age clusters only
if it starts with initial periods that are much longer than those measured in the youngest clusters
(dashed and dot-dashed curves in panels a and b in Fig.~\ref{fig:f1}).
However, the assumption that individual open clusters had their unique distributions of $P_0$ in the past, with very different statistics, does not seem
realistic because the same laws of physics had most likely shaped them during the star formation.
This conclusion is supported by the fact that the period distributions for solar-type stars in three of our youngest
clusters, ONC, NGC\,2264 and NGC\,2362, have high KS probabilities of having been drawn from the same real distribution (Fig.~\ref{fig:f2}).
Therefore, we will assume that the period distributions of stars in clusters of different ages $f(t,P)$ represent
an evolutionary sequence that started with the same distribution $f(0,P_0)$. Besides, we will consider the initial
distribution of disk-locking times to be flat and random in the interval $0 < \tau_{\rm d}\leq (\tau_{\rm d})_{\rm max}$
with the maximum value of $(\tau_{\rm d})_{\rm max} = 20$ Myr.
As a proxy for $f(0,P_0)$, we will take period distributions for the three aforementioned young clusters.
If the initial rotation of cluster stars was much slower than predicted by the ONC data then we should see
clusters arriving on the MS with very low rotation rates, e.g. systems with ages below 100 Myr should already
have a lot of slow rotators. Small samples from very young open clusters exhibit no such trend, but larger
sample sizes are required.

The assumptions that we have just made are not novel. For instance,
\cite{iea07} evolved the observed rotation rates of low-mass stars in
NGC\,2362 forward in time trying to reproduce some statistics of $\Omega_{\rm e}$ distributions
in older clusters. Like us, they employed a double-zone model. To compare the results of their computations with
observations, \cite{iea07} calculated the lower 25th and upper 90th percentiles of the observed distributions of
$\Omega_{\rm e}$ for solar-type stars in NGC\,2362, IC\,2391/IC\,2602, $\alpha$\,Per, M\,34, and
the Hyades. These statistics have been chosen to characterize the slowest and fastest rotators in the selected clusters.
Using our double-zone model with $\tau_{\rm c} = 1$ Myr, we confirm the conclusion made by \cite{iea07} that the rotational evolution of
the fastest rotators among solar-type stars can be simulated closely enough with short coupling times characteristic of SB rotators
(upper dashed curves in Fig.~\ref{fig:f8}). On the contrary, much longer coupling times of order
$\tau_{\rm c} = 50$\,--\,150 Myr and $\tau_{\rm c} = 100$\,--\,300 Myr for the mass bins
$0.9\leq M/M_\odot\leq 1.1$ and $0.7\leq M/M_\odot < 0.9$, respectively, as obtained in our computations
(solid curves in the same figure), are required to explain the rotational evolution of the slowest
rotators\footnote{\cite{iea07} reported much longer coupling times
because they did not recalibrate the wind parameter $K_{\rm w}$ for the slowest rotators.}.

From a theoretical standpoint, the inability of the SB rotation model, on the one hand, and the ability of
the DR model, on the other hand, to reproduce the location of the slowest rotators in the $\Omega_{\rm e}$-age plot
is caused by their, respectively, inappropriate and appropriate characteristics of mapping of $f(0,P_0)$ to $f(t,P)$
for long rotation periods. Apparently, the still unknown physical mechanism responsible for the internal angular momentum
transport in solar-type stars must have a property, which is most likely related to $\Omega_{\rm e}$, of making
rotation of the radiative core and convective envelope more and more decoupled as one goes from a faster to slower rotating star.
For example, the timescale for internal angular momentum transport by hydrodynamic mechanisms becomes shorter
as the rotation rate increases in this fashion (\citealt{pkd90}).
Assuming that a quantitative characteristic describing this property changes continuously with $\Omega_{\rm e}$ (or $P$),
this should result in qualitative differences between the true $\Omega_{\rm e}$ distributions for evolved clusters and those obtained
with the double-zone model for different values of $\tau_{\rm c}$. These qualitative differences can be put on a scale
by comparing the observed distributions of $\Omega_{\rm e}$ with the modeled ones using the Kolmogorov-Smirnov test.
This procedure is more rigorous than just matching the percentiles, because it uses statistical information
encoded in the entire distribution rather than only in a part of it.

Given the aforementioned robust results concerning the fastest solar-type rotators, the basic assumption of our
following statistical analysis is that the rotational evolution of stars with sufficiently short initial rotation periods
can be described by the double-zone model with $\tau_{\rm c} = 1$ Myr, whereas that of slower rotating stars
needs $\tau_{\rm c}\gg 1$ Myr. In principle, we ought to introduce some monotonically increasing function
$\tau_{\rm c}(P_0)$ into our double-zone model that would produce a smooth transition from short-period SB rotators
to stars with progressing DR that had longer initial periods. However, given the simplicity of the double-zone model,
such approach looks over-complicated. Therefore, we have decided to employ a simple step function
\bea
\tau_{\rm c}(P_0) = \left\{ 
  \begin{array}{l l}
    1 \mbox{ Myr,} & \quad \mbox{if } P_0\le P_{\rm c},\\
      \tau_{\rm c}\gg 1 \mbox{ Myr,} & \quad \mbox{if } P_0 > P_{\rm c},\\
  \end{array} \right.
\label{eq:ptau}
\eea
where $P_{\rm c}$ is a critical period.

The main objective of our statistical analysis of open-cluster rotation period data for solar-type stars
is to get estimates of the combination of parameters $(P_{\rm c},\,\tau_{\rm c})$
from (\ref{eq:ptau}) that give the highest KS probabilities of the hypothesis that an observed $\Omega_{\rm e}$ distribution for an open cluster of age $t$
and our theoretical $\Omega_{\rm e}$ distribution computed for the same age $t$
have been drawn from the same real distribution. It is assumed that the initial period distribution
$f(0,P_0)$ is provided by the youngest clusters from our data compilation: ONC, NGC\,2362, or NGC\,2264 (Table~\ref{tab:tab1}). 
To do the analysis, we have performed extensive double-zone model computations densely covering relevant regions (squares) of
the $((\tau_{\rm d})_{\rm max},\,\tau_{\rm c})$ parameter space for a number of $P_{\rm c}$ values.
The resulting $P_{\rm KS}$ contours are plotted in Fig.~\ref{fig:f10} (the mapping of ONC to M\,35 for the mass bin
centered at $M = 1.0\,M_\odot$), Fig.~\ref{fig:f11} (NGC\,2362 to M\,50 for $M = 1.0\,M_\odot$), Fig.~\ref{fig:f12}
(the mapping of NGC\,2362 to M\,34 for the mass bin centered at $M = 0.8\,M_\odot$), and Fig.~\ref{fig:f13}
(NGC\,2362 to M\,50 for $M = 0.8\,M_\odot$).

The $P_{\rm KS}$ contour patterns revealed in Figs.~\ref{fig:f10}, \ref{fig:f11}, and \ref{fig:f12}
clearly show a decrease of the most probable value of the coupling time with an increase of the disk-locking time,
as expected. Indeed, a star whose pre-MS rotation rate was being kept constant, as a result of
its interaction with a circumstellar disk, for a longer time would arrive at the ZAMS having a smaller amount of angular momentum stored
in the radiative core. Therefore, even a shorter coupling time will not lead to a fast rotation of its convective envelope
simply because there is not much of angular momentum left in the core to be transported to the envelope.

From Fig.~\ref{fig:f2} and Table~\ref{tab:tab1} we can see that there is
a large number of stars in the youngest clusters with periods $P_0 < 12$ days. Yet the $P_{\rm KS}$ contour patterns
in Figs.~\ref{fig:f10}, \ref{fig:f11}, and \ref{fig:f12} first become apparent, as $P_{\rm c}$ increases,
and then get dissolved well before the critical period reaches the value of 12 days.
This means that the double zone model cannot reproduce the $\Omega_{\rm e}$ distributions
in evolved clusters under the assumption that {\it all} stars are SB rotators. There is always a group of stars with
rotation of the radiative core decoupled from that of the convective envelope. For the mass bin centered at $1.0\,M_\odot$, 
a lower limit for the coupling time can be estimated as $\tau_{\rm c} \ga 30$ Myr
for $(\tau_{\rm d})_{\rm max} = 20$ Myr, and $\tau_{\rm c} \ga 55$ Myr for $(\tau_{\rm d})_{\rm max} = 10$ Myr (Fig.~\ref{fig:f10}).
Fig.~\ref{fig:f11} gives nearly 20 Myr longer coupling times.
Probable values of the parameter $P_{\rm c}$ that divides stars into SB rotators and objects possessing DR range from    
2.5 to 5.5 days for these two mappings. However, what is more important for us here is that, in all considered cases, we can completely rule out
the pure SB rotation evolution ($P_{\rm c} = \infty$) as a solution. Consequently, Spruit's mechanism or any other
physical mechanism that can only produce nearly SB rotation should be rejected as a prescription for rotational evolution of
solar-type stars. The true model should have a parameter (presumably related to $\Omega_{\rm e}$) that allows it to tune up
its rotational evolution so that the latter would resemble the rotational evolution of the double-zone model with
the coupling time changing from $\tau_{\rm c} = 1$ Myr to $\tau_{\rm c} \approx 55$ Myr.
For stars with masses in the interval $0.7\leq M/M_\odot < 0.9$, we estimate much longer coupling times:
$\tau_{\rm c} \ga 150$ Myr for $(\tau_{\rm d})_{\rm max} = 20$ Myr, and $\tau_{\rm c} \ga 200$ Myr for $(\tau_{\rm d})_{\rm max} = 10$ Myr 
(Fig.~\ref{fig:f12} and Fig.~\ref{fig:f13}). This result can be anticipated from Fig.~\ref{fig:f8} where
we compare the lower 10th percentiles with our DR double-zone model computations.

As an additional statistical exercise aiming to demonstrate the failure of SB rotation to serve as a unique solution
for all solar-type stars, we have evolved
the observed period distributions for M\,35 and M\,50 backward in time to zero age, assuming the short
coupling time of 1 Myr for all stars, to see how their initial period distributions might resemble
those of Orion and NGC\,2264, respectively. To solve this inverse problem,
we have first projected (using our double-zone model with $\tau_{\rm c} = 1$ Myr)
a sufficiently wide rectangular domain of the $(\log P_0,\,\log\tau_{\rm d})$ parameter space into 
the $\log P$ space at the ages of M\,35 and M\,50. The projections have turned out to look like
perfect planes. Contour lines for the M\,35 projection are plotted in upper panel in Fig.~\ref{fig:f9}.
It is seen that the backward solution is not unique. In fact, every value of $\log P(t)$ at an older age $t$ is mapped    
into a diagonal line segment 
\bea
\log P_0 \approx a\log\tau_{\rm d}+b
\label{eq:diag}
\eea
back at $t=0$. To assign a unique value of the
initial period for $\log P(t)$, we have decided to randomly select $\tau_{\rm d}$ from an interval
restricted by the end points of the diagonal (\ref{eq:diag}). Thus computed ``initial'' rotation periods
for the M\,35 and M\,50 stars are compared with the Orion and NGC\,2264 period distributions
in Fig.~\ref{fig:f9} (middle and bottom panels). 

We see that the hypothesis that all solar-type stars,
independently of their rotation periods, evolved like SB rotators would require that
M\,35 and M\,50 had initially contained much larger fractions of slow rotators
compared to Orion and NGC\,2264. Because this is unlikely to be the case,
we should reject any prescription for the internal transport of angular momentum that 
enforces SB rotation in a solar-type star without respect to how slowly it rotates. In particular,
the original Tayler-Spruit dynamo cannot be considered as a relevant mechanism for angular 
momentum redistribution in solar-type stars because it always enforces nearly SB rotation in both
fastest and slowest rotators.

\section{Discussion and Conclusions}

\begin{figure}
\epsfxsize=14cm
\epsffile [60 270 480 695] {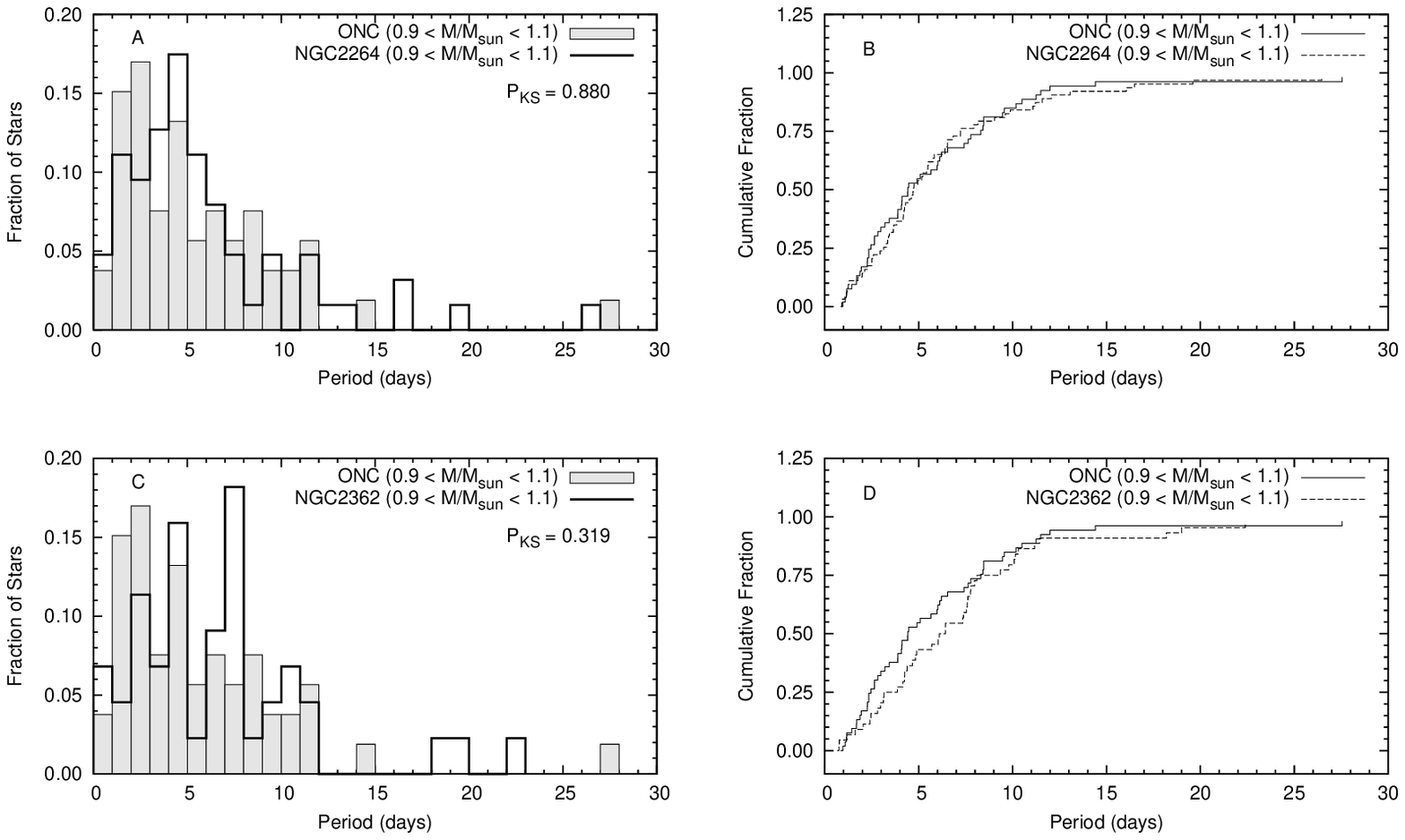}
\caption{Panels a and c: Period distributions for solar-type stars in the Orion cluster (shaded histograms)
         are compared to those in NGC\,2264 and NGC\,2362 (thick solid curves). The Kolmogorov-Smirnov test
         gives high probabilities ($P_{\rm KS} \gg 0.05$) that the corresponding pairs of distributions have been drawn
         from the same real distribution. Panels b and d: Cumulative distribution functions
         for the two pairs of clusters.}
\label{fig:f2}
\end{figure}

\begin{figure}
\epsfxsize=13cm
\epsffile [60 180 480 695] {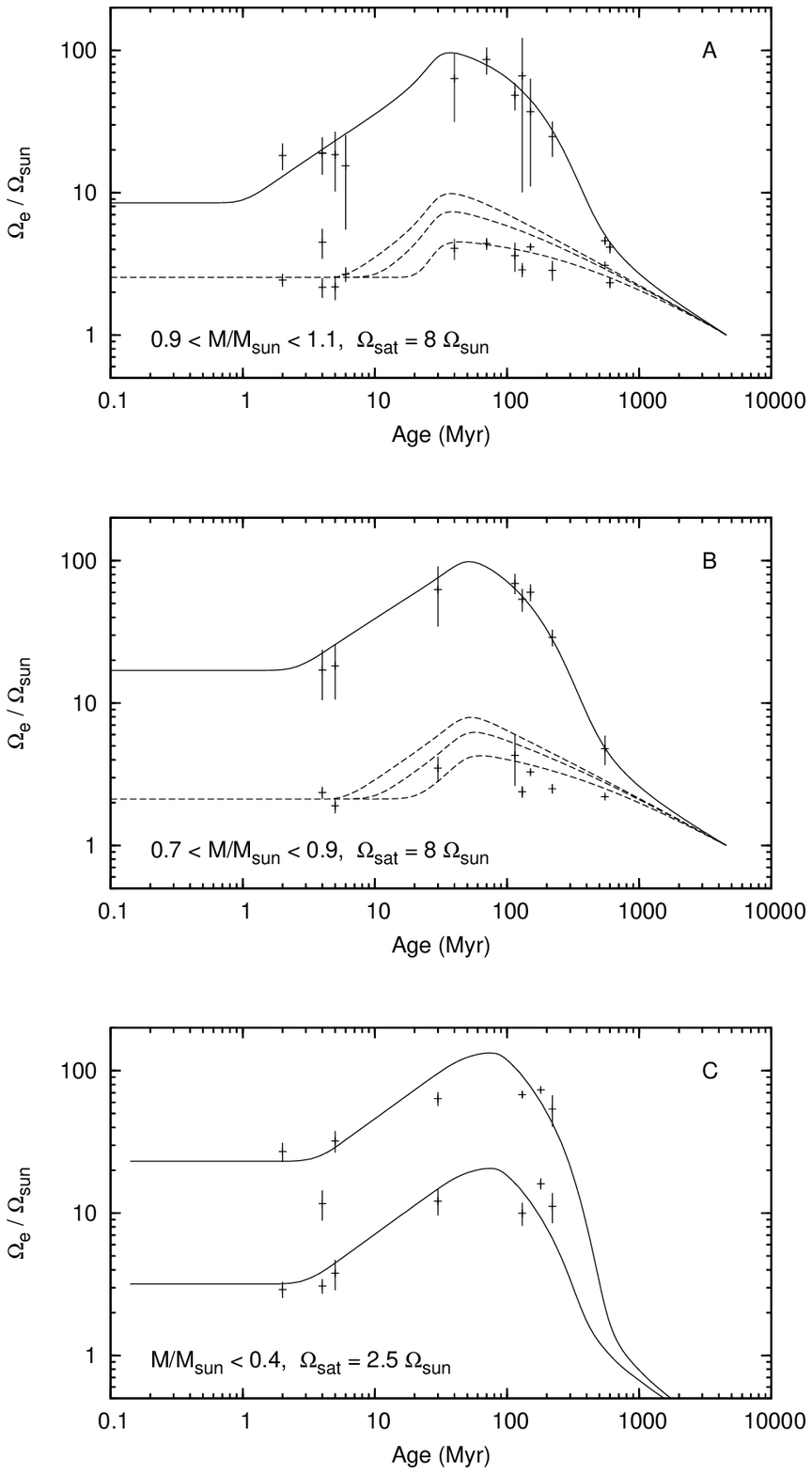}
\caption{Upper 90th and lower 10th percentiles of $\Omega_{\rm e} = 2\pi/P$ distributions (crosses) for
         all of our compiled data samples (Table~\ref{tab:tab1}). The 90th percentiles are used
         to adjust the parameter $\Omega_{\rm sat}$ (shown in each panel) by fitting them to the SB rotation
         evolution (solid curves). The 10th percentiles for the fully convective stars are also fitted
         well by a SB rotation evolution curve (panel c). The percentile vertical errorbars were evaluated through
         bootstrap simulations.
         } 
\label{fig:f3}
\end{figure}

\begin{figure}
\epsfxsize=13cm
\epsffile [10 10 450 400] {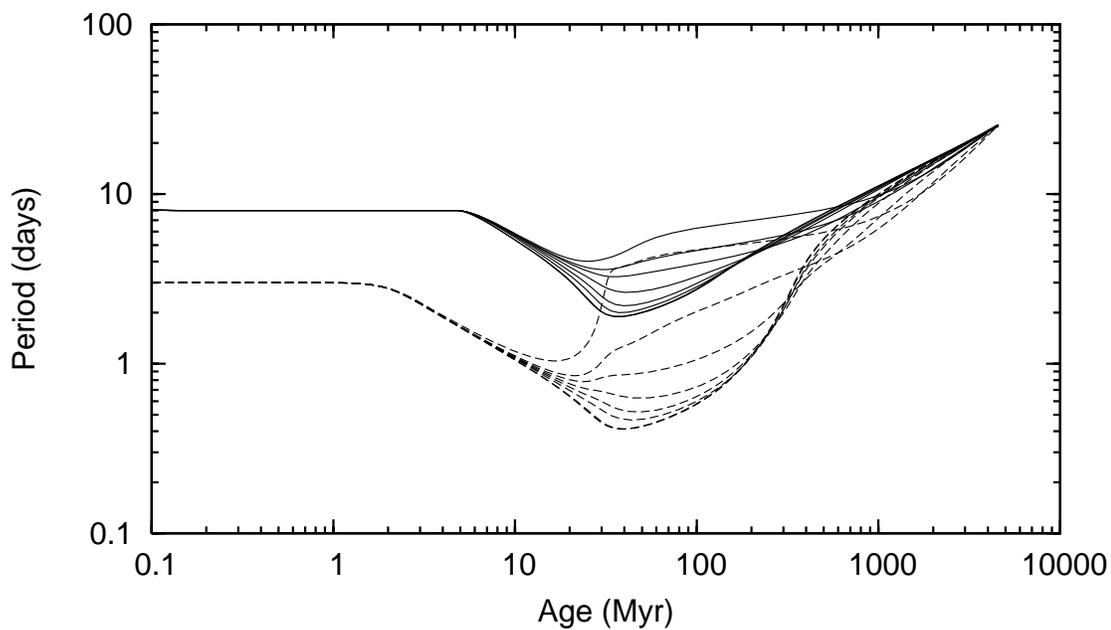}
\caption{Rotation period evolution of the Sun computed using the constant viscosity model
         for the initial conditions $(\tau_{\rm d},\,P_0) = (6\,\mbox{Myr},\,8\,\mbox{days})$
         (solid curves), and $(\tau_{\rm d},\,P_0) = (2\,\mbox{Myr},\,3\,\mbox{days})$
         (dashed curves), and for the values of $\nu_0 = 2.5\times 10^4,\ 5\times 10^4,\ 10^5,\ 
         2.5\times 10^5,\ 5\times 10^5,\ 10^6,\ 10^7$, and $10^8$ cm$^2$\,s$^{-1}$
         (from upper to lower curve for either combination of initial conditions).
         } 
\label{fig:f4}
\end{figure}

\begin{figure}
\epsfxsize=13cm
\epsffile [10 10 450 400] {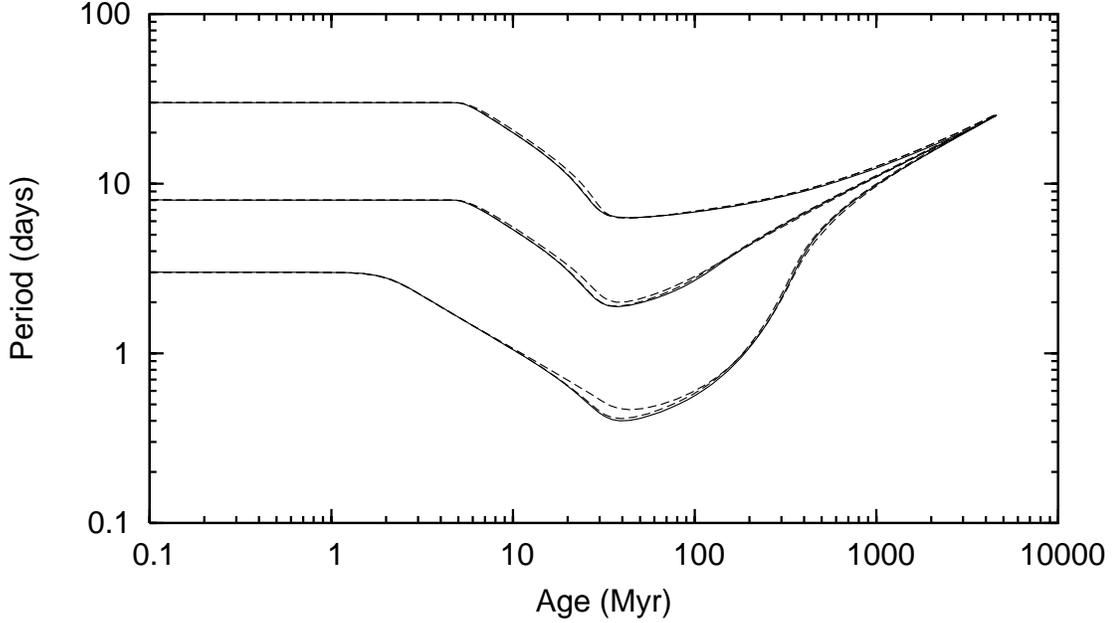}
\caption{Rotation period evolution computed for the initial conditions
         $(\tau_{\rm d},\,P_0) = (6\,\mbox{Myr},\,8\,\mbox{days})$,
         $(6\,\mbox{Myr},\,30\,\mbox{days})$, and $(2\,\mbox{Myr},\,3\,\mbox{days})$
         using the original prescription for the Tayler-Spruit dynamo
         (solid curves) and the constant viscosity model with
         $\nu_0 = 10^6$, and $10^7$ cm$^2$\,s$^{-1}$ (dashed curves).
         The dashed curves for $\nu_0 = 10^7$ cm$^2$\,s$^{-1}$ almost coincide
         with the solid curves. All computations have been performed
         with $M = 1\,M_\odot$ and $\Omega_{\rm sat} = 8\,\Omega_\odot$.
         } 
\label{fig:f5}
\end{figure}

One of the main conclusions made in this paper is that the observed rotation period distributions of
low-mass stars in open clusters do not seem to support the hypothesis that all solar-type stars
evolve as SB rotators. Instead, statistical analyses of these data show that
only the fastest rotators among solar-type stars can be considered to possess SB rotation during their entire evolution, whereas
their slowly rotating counterparts are most likely to manifest internal DR between the ages of $\sim$\,30 Myr and several hundred Myr. 
This is not a new result though because, e.g., \cite{iea07} came to a similar conclusion. A novelty of our work is that we have used
the entire period distributions of solar-type stars in a number of open clusters of different ages, not just some of their statistics, 
and extensive Monte-Carlo simulations to put this conclusion on a more rigorous quantitative basis.
This seems to be quite a reasonable approach to the solution of the problem, given that it has
a large number of poorly constrained parameters. In particular, we have found that a star
with $M = 1.0\pm 0.1\,M_\odot$ that starts its rotational evolution with a period $P_0 \ga 2$\,--\,4 days
should have rotation of its convective envelope and radiative core coupled on a timescale of
order $\tau_{\rm c} = 55\pm 25$ Myr, where the systematic uncertainty of this estimate takes into account
the anticipated decrease of the coupling time with an increase of the disk-locking time.
For a slightly less massive star with $M = 0.8\pm 0.1\,M_\odot$, the coupling time increases to $\tau_{\rm c} = 175\pm 25$ Myr.
Given that the initial period distributions, those for the youngest open clusters, have rather densely occupied
bins up to $P_0 \approx 12$ days (Fig.~\ref{fig:f2}), it turns out that quite large fractions of solar-type stars
(up to 50\%) should go through a phase of DR evolution.

It is important to note that solar-type stars in open clusters older than a few hundred Myr are not suitable for a type of
statistical analysis employed by us. The problem with the older clusters is that rotation periods of
solar-type stars in them have already converged too close to each other, all aiming eventually to approach the solar rotation.
Therefore, period distributions for the older clusters do not allow to make an unambiguous conclusion about
the rotational evolution of solar type stars. To illustrate this, we have evolved the NGC\,2264 period distribution of
stars with $M = 1.0\pm 0.1\,M_\odot$ to a period distribution at $t = 550$ Myr corresponding to the age of M\,37 (Fig.~\ref{fig:f14}).
We see that at this old age the period distribution mapping admits two branches of solutions, one with DR and another with SB rotation.
This bimodality is caused by the very narrow range of the mapping at old ages which finally degenerates into a point
at the solar age.

The second novelty of our investigation is that we have related the coupling time from the double-zone model
to its corresponding value of the constant viscosity from our full stellar evolutionary model (Fig.~\ref{fig:f7}).
In particular, the minimum coupling time of 30 Myr obtained in our analysis of DR of solar-type stars
roughly corresponds to $\nu_0 = 7.5\times 10^4$ cm$^2$\,s$^{-1}$, whereas the longest coupling time of
$\sim$80 Myr for stars with $M = 1.0\pm 0.1\,M_\odot$ implies that in some of them the internal transport of
angular momentum can be as slow as that modeled with a viscosity $\nu_0\approx 3\times 10^4$ cm$^2$\,s$^{-1}$.

Finally, we have shown that original Spruit's prescription always enforces SB rotation in a solar-type star,
no matter how slowly it rotates. Therefore, it cannot be accepted as a physical mechanism for the internal angular
momentum redistribution in radiative cores of solar-type stars many of which are indirectly proved to possess a degree of DR.

\begin{figure}
\epsfxsize=12cm
\epsffile [50 180 400 650] {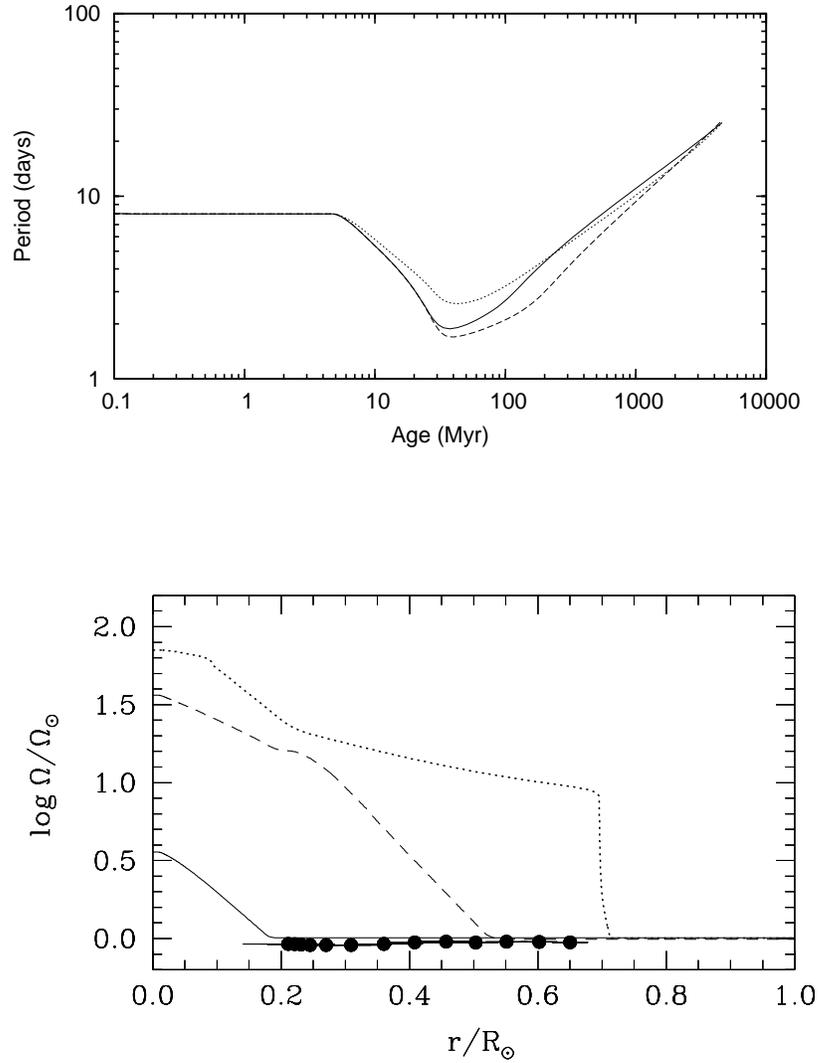}
\caption{Rotation period evolution (upper panel) and final rotation profiles
         in the solar models (bottom panel) computed using Spruit's original (solid curves) and revised
         prescription (dashed curves). Dotted curves represent a model with no
         internal angular momentum transport. Filled circles in the bottom panel are
         helioseismic data from \cite{cea03}.
         } 
\label{fig:f6}
\end{figure}

\begin{figure}
\epsfxsize=12cm
\epsffile [50 180 400 650] {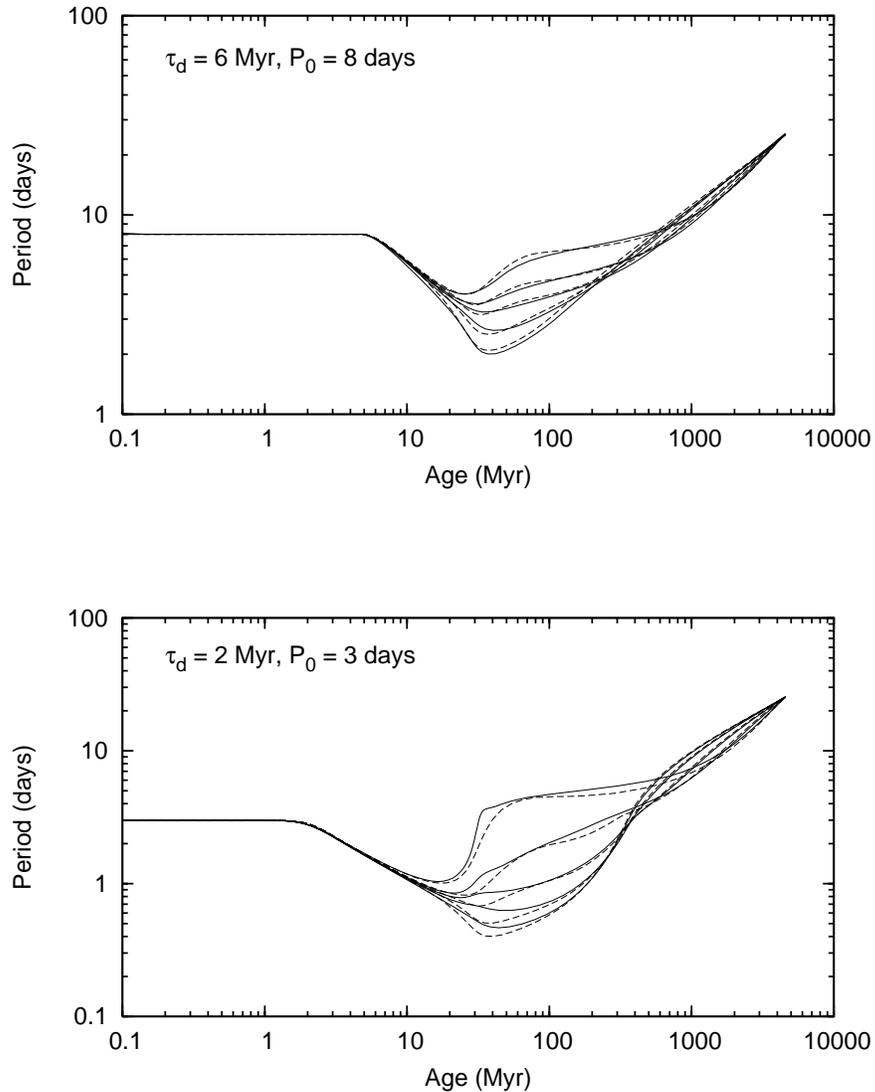}
\caption{Comparison of rotation period evolution of the Sun computed using the constant
         viscosity (solid curves) and double-zone model (dashed curves).
         The corresponding pairs of the viscosity and core/envelope coupling time adjusted
         for the models' $P$-age relations to resemble one another as closely
         as possible are $(\nu_0,\,\tau_{\rm c}) =
         (2.5\times 10^4,\,90),\ (5\times 10^4,\,40),\ (10^5,\,20),\
         (2.5\times 10^5,\,7)$, and $(10^6,\,1)$, where $\nu_0$ is given in
         cm$^2$\,s$^{-1}$, and $\tau_{\rm c}$ in Myr.
         } 
\label{fig:f7}
\end{figure}

\begin{figure}
\epsfxsize=12cm
\epsffile [60 180 480 695] {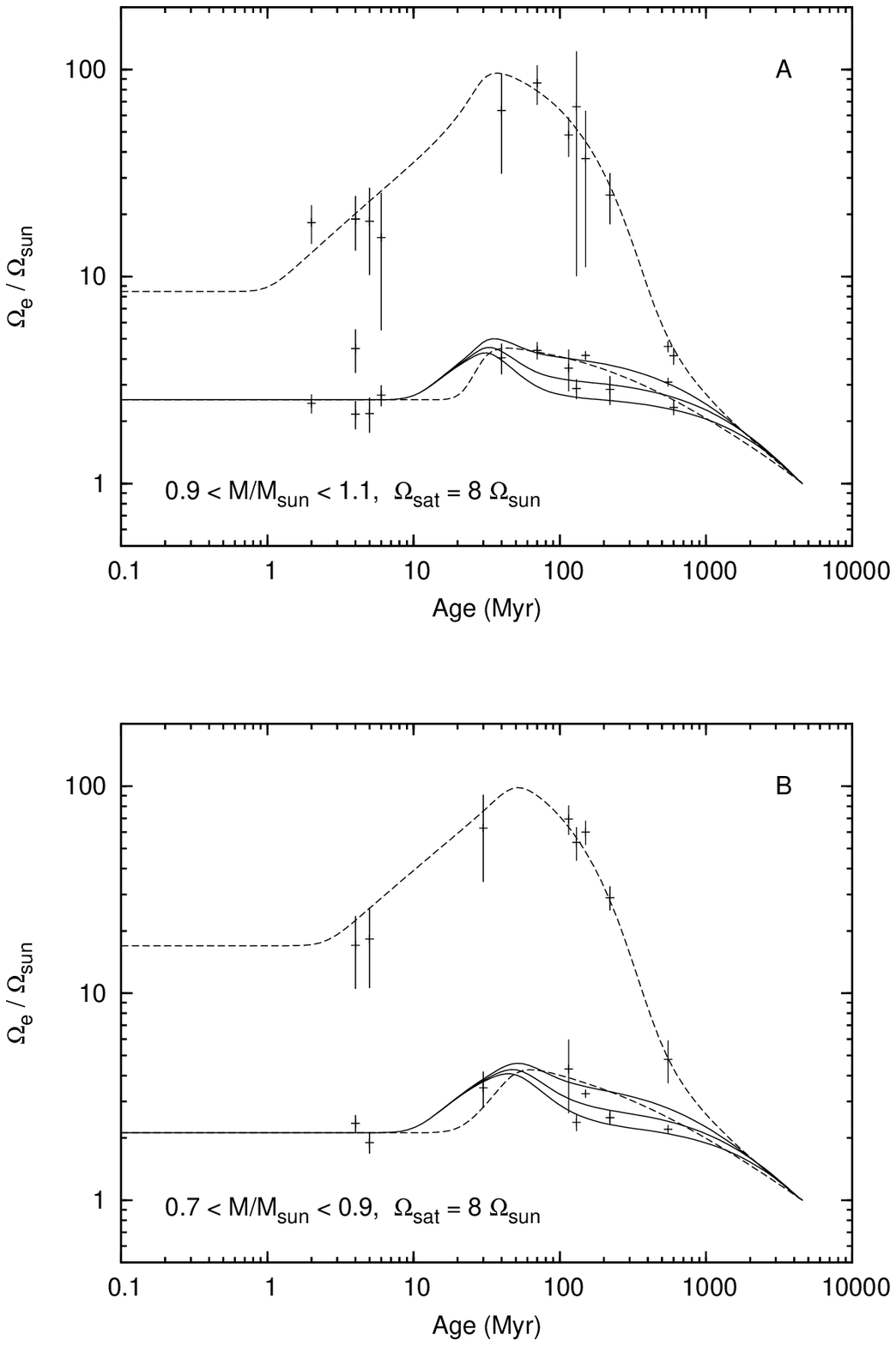}
\caption{Upper 90th and lower 10th percentiles (crosses) of $\Omega_{\rm e}$ distributions
         for all of our compiled data samples of open cluster solar-type stars are compared with $\Omega_{\rm e}$-age relations
         computed using the double-zone model. Dashed curves
         correspond to $\tau_{\rm c} = 1$\,Myr (SB rotation), solid curves (from upper to lower) to DR with
         $\tau_{\rm c} = 50$, 100 and 150 Myr (panel a), and $\tau_{\rm c} = 100$, 200 and 300 Myr (panel b).
        } 
\label{fig:f8}
\end{figure}

\begin{figure}
\epsfxsize=14cm
\epsffile [60 220 480 695] {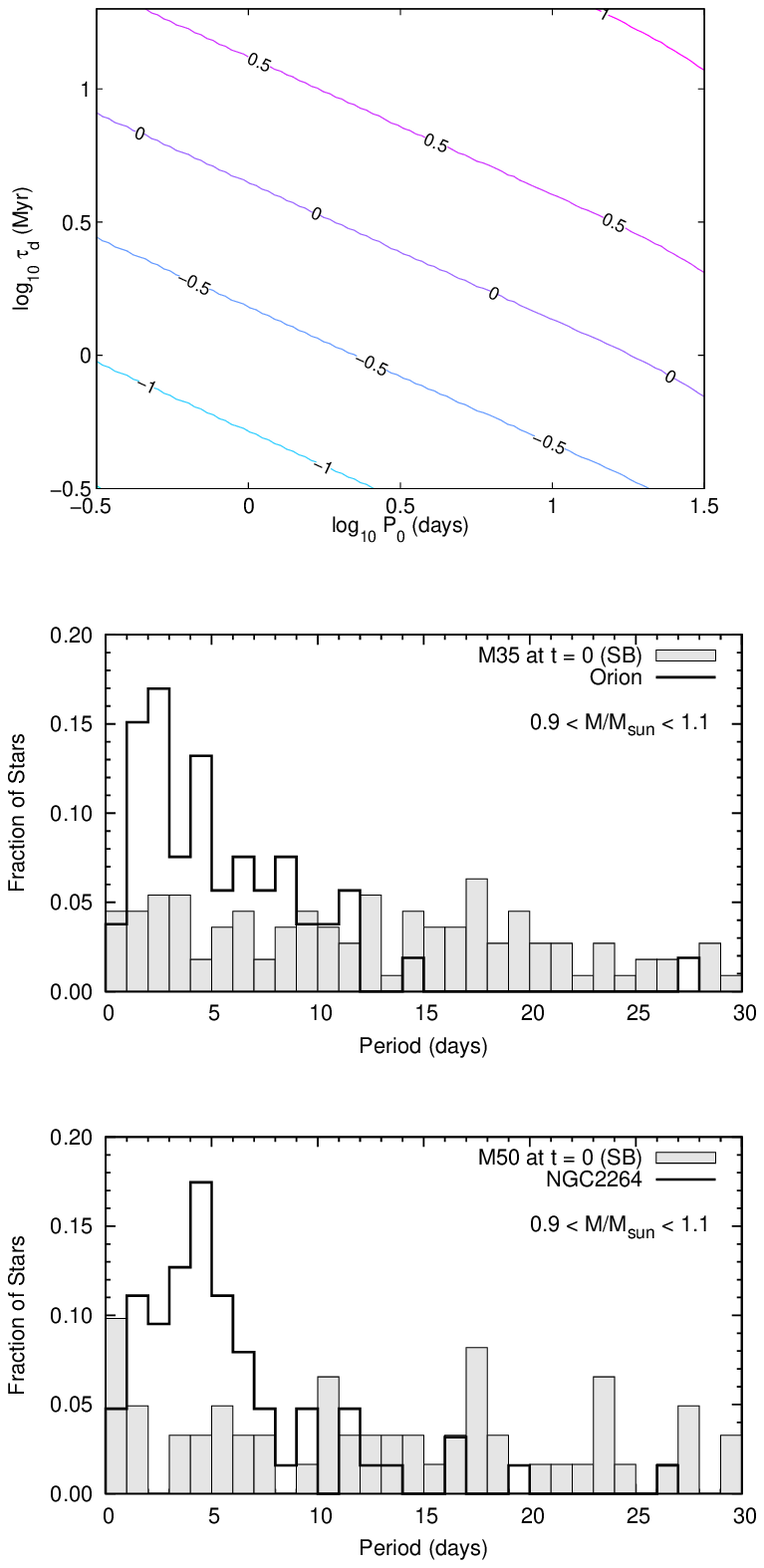}
\caption{Upper panel ---contour lines in a domain of initial conditions $(\log P_0,\,\log\tau_{\rm d})$
         that are projected to specified values of $\log P$ (shown on the lines)
         at the age of M\,35 (150 Myr) for SB rotational
         evolution computed using the double-zone model with $\tau_{\rm c} = 1$ Myr.
         These lines are used to restore the initial period distribution for solar-type stars
         in M\,35 (shaded histogram in middle panel) and to compare it with the Orion period distribution
         (thick solid curve). Bottom panel shows the corresponding comparison between M\,50 and NGC\,2264.
         } 
\label{fig:f9}
\end{figure}

\begin{figure}
\epsfxsize=14cm
\epsffile [60 260 480 695] {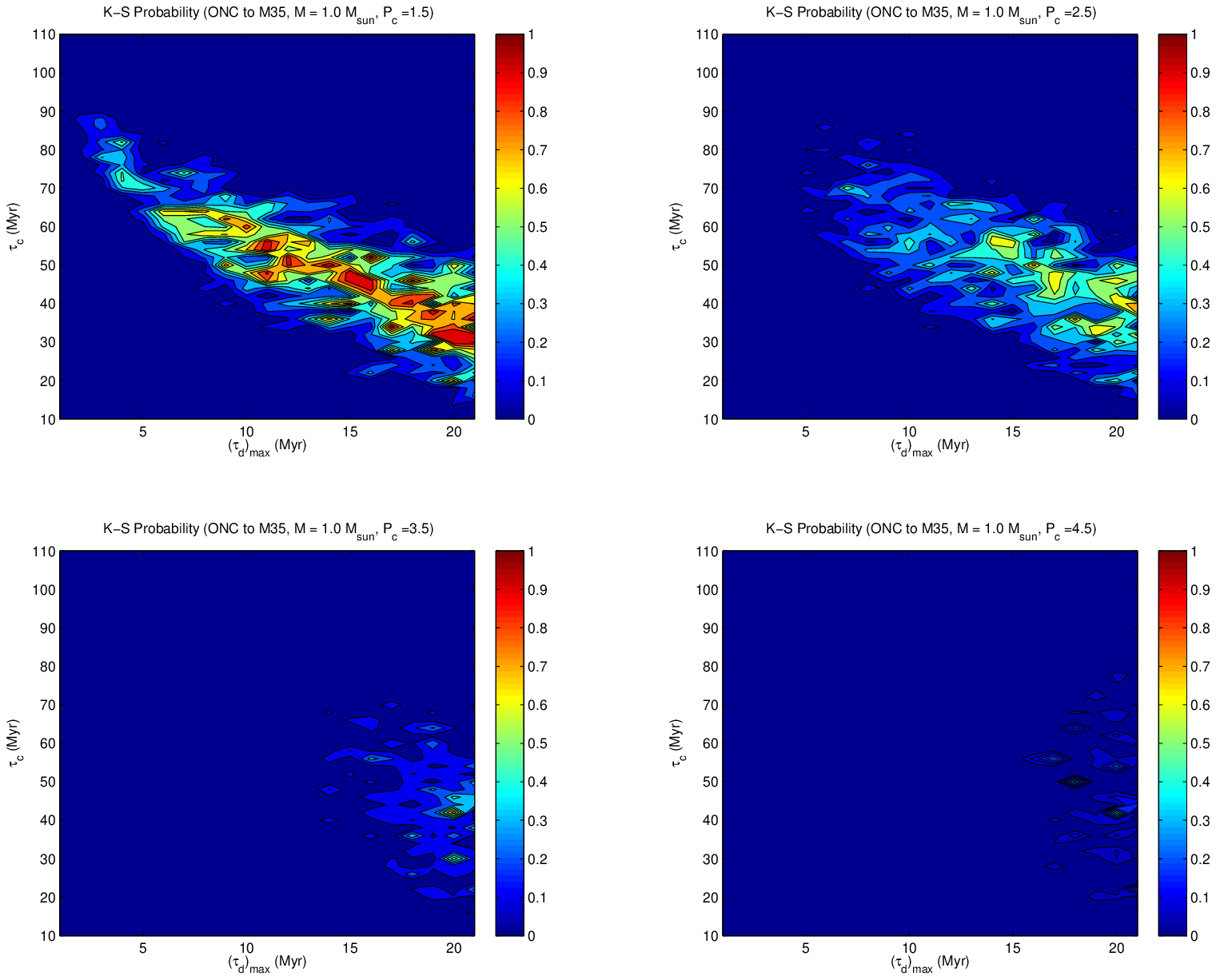}
\caption{Contours of the Kolmogorov-Smirnov probability of the hypothesis that the observed and modeled
         period distributions of M\,35 stars with $M = 1.0\pm 0.1\,M_\odot$ have been drawn from
         the same real distribution. The initial periods are taken from the ONC cluster.
         The double-zone model with the dependence (\ref{eq:ptau}) of $\tau_{\rm c}$ on $P_0$
         has been employed. Simulations have been done for the shown intervals of the parameters
         $\tau_{\rm c}$ and $(\tau_{\rm d})_{\rm max}$, and for the values of $P_{\rm c}$ specified
         in parenthesis atop each panel.
         } 
\label{fig:f10}
\end{figure}

\begin{figure}
\epsfxsize=11cm
\epsffile [60 180 480 695] {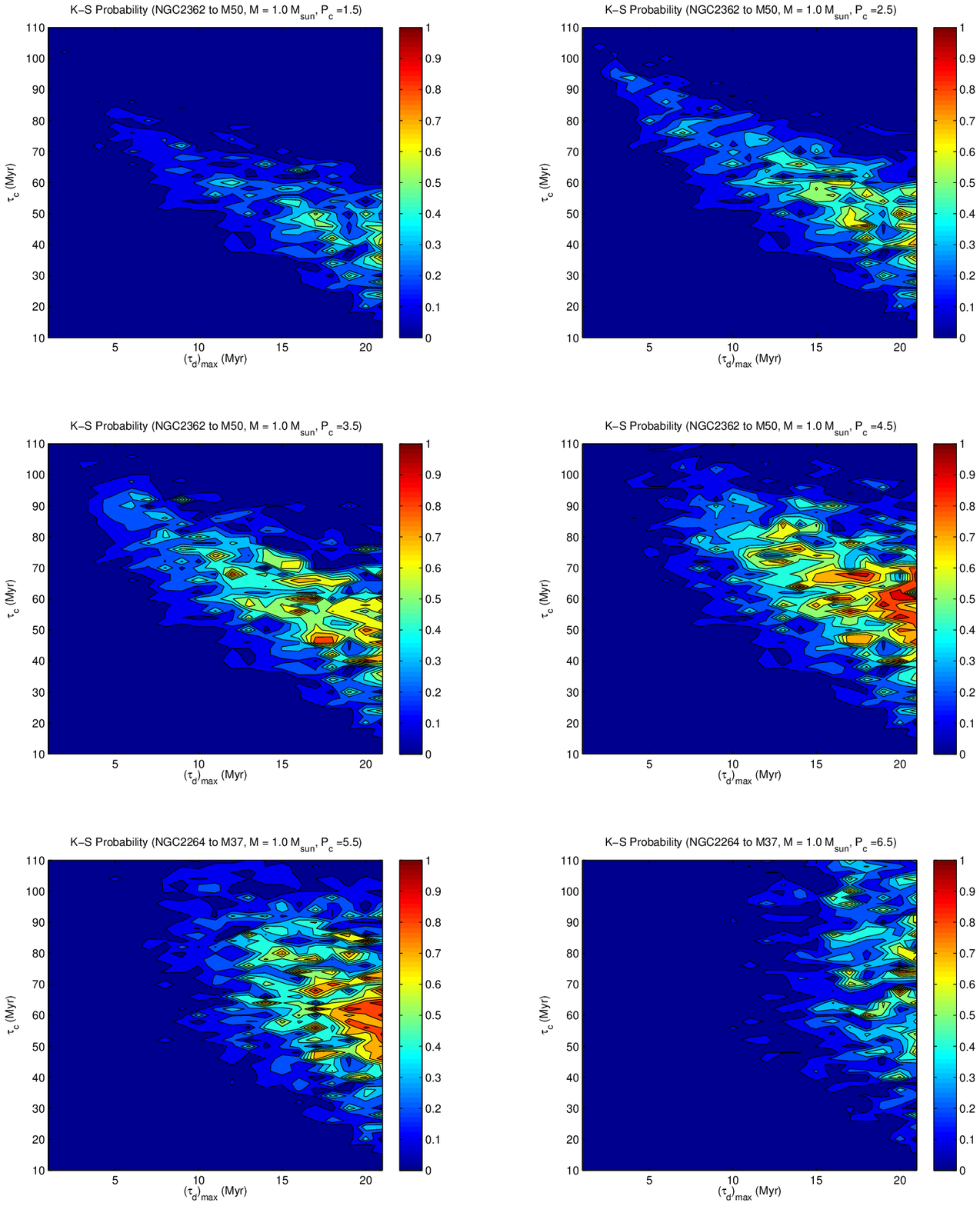}
\caption{Contours of the Kolmogorov-Smirnov probability of the hypothesis that the observed and modeled
         period distributions of M\,50 stars with $M = 1.0\pm 0.1\,M_\odot$ have been drawn from
         the same real distribution. The initial periods are taken from the NGC\,2362 cluster.
         The double-zone model with the dependence (\ref{eq:ptau}) of $\tau_{\rm c}$ on $P_0$
         has been employed. Simulations have been done for the shown intervals of the parameters
         $\tau_{\rm c}$ and $(\tau_{\rm d})_{\rm max}$, and for the values of $P_{\rm c}$ specified
         in parenthesis atop each panel.
         } 
\label{fig:f11}
\end{figure}

\begin{figure}
\epsfxsize=11cm
\epsffile [60 80 480 745] {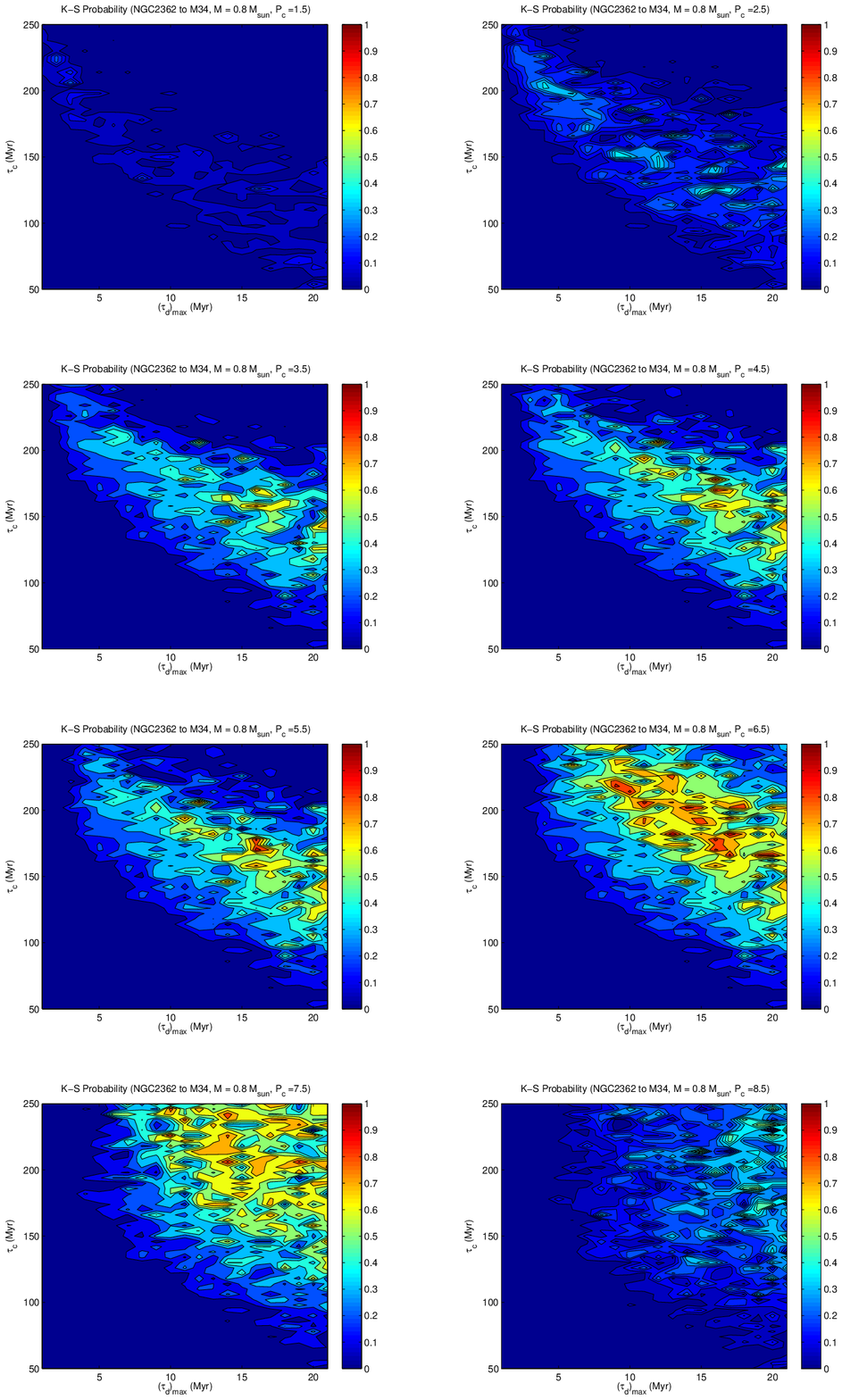}
\caption{Contours of the Kolmogorov-Smirnov probability of the hypothesis that the observed and modeled
         period distributions of M\,34 stars with $M = 0.8\pm 0.1\,M_\odot$ have been drawn from
         the same real distribution. The initial periods are taken from the NGC\,2362 cluster.
         The double-zone model with the dependence (\ref{eq:ptau}) of $\tau_{\rm c}$ on $P_0$
         has been employed. Simulations have been done for the shown intervals of the parameters
         $\tau_{\rm c}$ and $(\tau_{\rm d})_{\rm max}$, and for the values of $P_{\rm c}$ specified
         in parenthesis atop each panel.
         } 
\label{fig:f12}
\end{figure}

\begin{figure}
\epsfxsize=11cm
\epsffile [60 80 480 745] {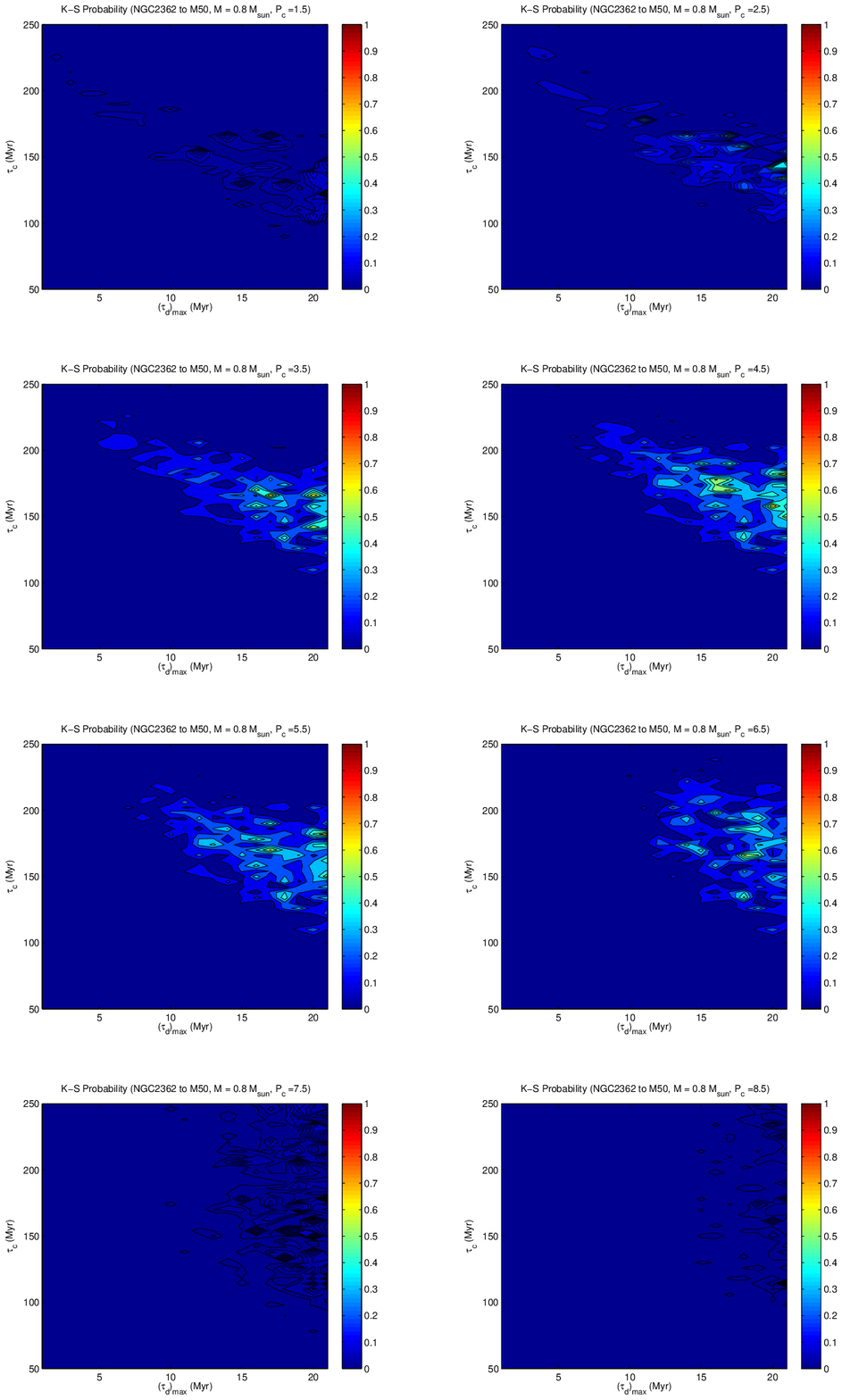}
\caption{Contours of the Kolmogorov-Smirnov probability of the hypothesis that the observed and modeled
         period distributions of M\,50 stars with $M = 0.8\pm 0.1\,M_\odot$ have been drawn from
         the same real distribution. The initial periods are taken from the NGC\,2362 cluster.
         The double-zone model with the dependence (\ref{eq:ptau}) of $\tau_{\rm c}$ on $P_0$
         has been employed. Simulations have been done for the shown intervals of the parameters
         $\tau_{\rm c}$ and $(\tau_{\rm d})_{\rm max}$, and for the values of $P_{\rm c}$ specified
         in parenthesis atop each panel.
         } 
\label{fig:f13}
\end{figure}

\begin{figure}
\epsfxsize=11cm
\epsffile [60 180 480 695] {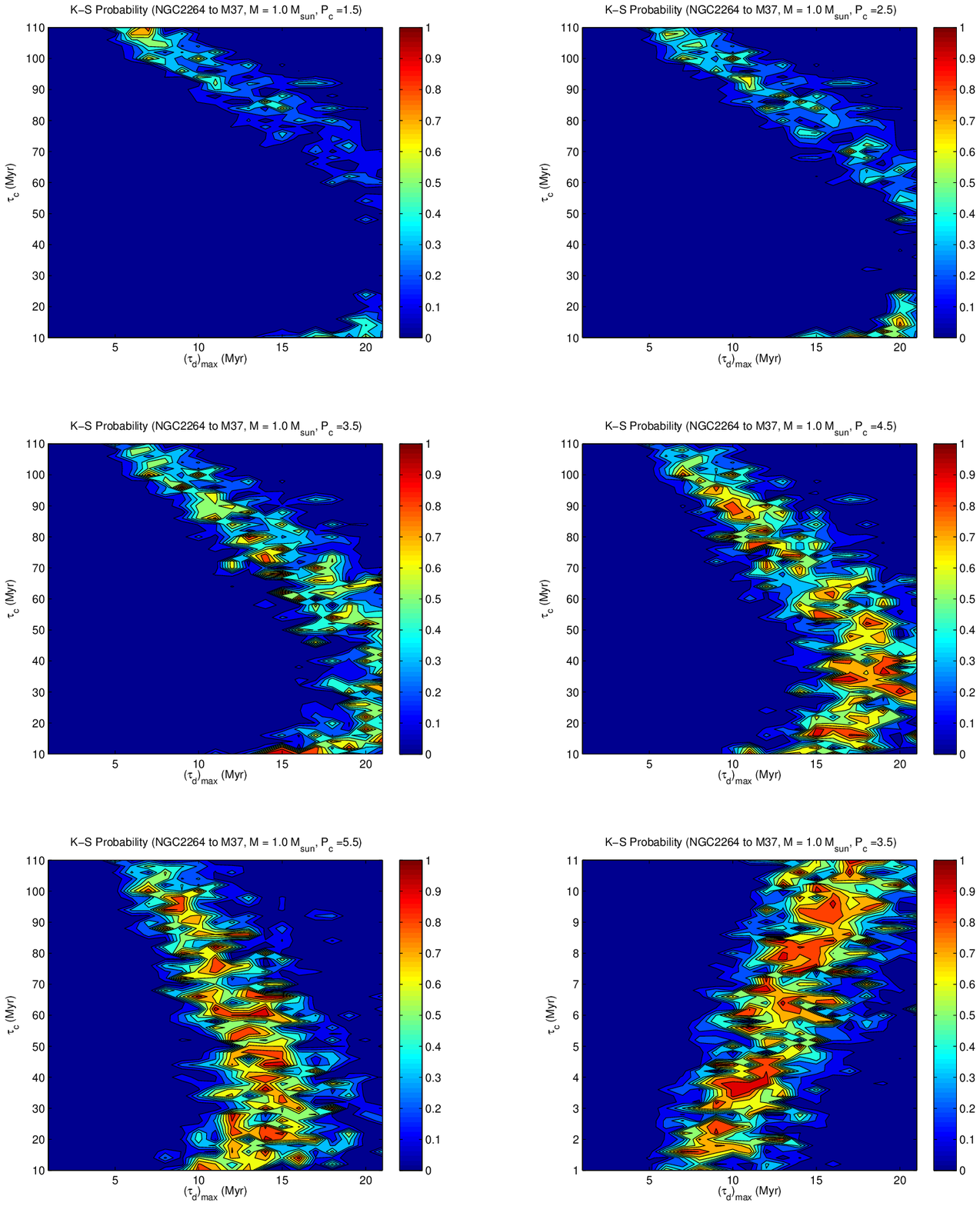}
\caption{Contours of the Kolmogorov-Smirnov probability of the hypothesis that the observed and modeled
         period distributions of M\,37 stars with $M = 1.0\pm 0.1\,M_\odot$ have been drawn from
         the same real distribution. The initial periods are taken from the NGC\,2264 cluster.
         The double-zone model with the dependence (\ref{eq:ptau}) of $\tau_{\rm c}$ on $P_0$
         has been employed. Simulations have been done for the shown intervals of the parameters
         $\tau_{\rm c}$ and $(\tau_{\rm d})_{\rm max}$, and for the values of $P_{\rm c}$ specified
         in parenthesis atop each panel. Because of the convergence to
         the solar rotation, there is an ambiguity in the rotational evolution mode (SB or DR) for
         this 550 Myr old cluster.
         } 
\label{fig:f14}
\end{figure}

We have also found that the revised prescription for the Tayler-Spruit dynamo (\citealt{dp07})
results in the rotational evolution superficially resembling that obtained using the original prescription,
although the former leaves a large differentially rotating radiative core in the present-day solar model
(Fig.~\ref{fig:f6}). Hence, the revised prescription appears to be in conflict both  with
the rotation period data for solar-type stars 
in open clusters (like the original prescription) and with the helioseismic data (unlike the original prescription).
It is worth seeing if other possible angular momentum transport  mechanisms
can assist the revised prescription in bringing the rotating solar model closer to
observations. To test this idea, we have supplemented the effective magnetic viscosity (\ref{eq:nueminrev})
with viscosities arising from the secular shear instability (denoted by the subscript ``ss'' below),
Goldreich-Schubert-Fricke instability (GSF), as well as with the molecular viscosity (mol) and a viscosity 
$\nu_{\rm mc}=|rU(r)|/5$
that approximately describes the transport of angular momentum by meridional circulation as a diffusion process:
\bea
\nu = (\nu_{\rm e})_{\rm min}^{(\ref{eq:nueminrev})} + \nu_{\rm ss} + \nu_{\rm GSF} + \nu_{\rm mc} + \nu_{\rm mol}.
\label{eq:allvisc}
\eea
To calculate the quantities $\nu_{\rm ss}$, $\nu_{\rm GSF}$, and the meridional circulation velocity $U(r)$,
we have used the corresponding equations from Appendix to the paper of \cite{chpt05}. Note that their expression for $U(r)$      
neglects terms depending on the $\mu$-gradient and its derivatives. Therefore, the meridional circulation
in this approximation is expected to penetrate deeper into the radiative core than in the case of its
fully consistent implementation originally proposed by \cite{mz98} that had later been applied to
study the rotational mixing in solar-metallicity MS stars with $M\geq 1.35\,M_\odot$ by \cite{pea03}.

\begin{figure}
\epsfxsize=10cm
\epsffile [70 240 420 700] {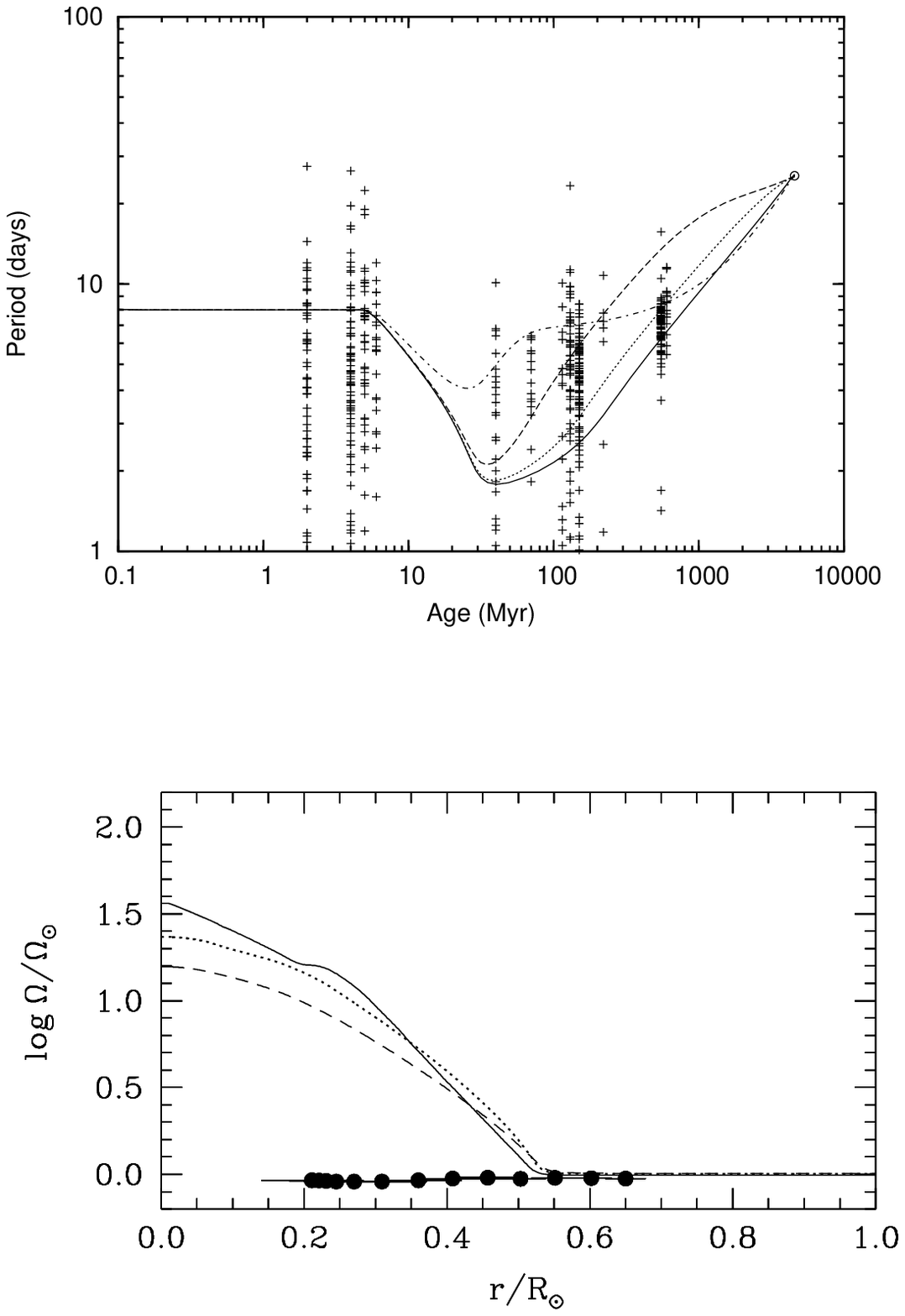}
\caption{Rotation period evolution and final rotation profiles in the solar models
         computed using Spruit's revised prescription (solid curves) and the combined
         viscosity (\ref{eq:allvisc}) (dotted curves). Dashed curves represent results
         obtained with the combined viscosity in which the term $\nu_{\rm GSF}$ has been
         multiplied by a factor of 10. For comparison, dot-dashed curve in upper panel
         shows the rotational evolution computed using the double-zone model with
         $\tau_{\rm c} = 100$ Myr (a model with DR).
         } 
\label{fig:f15}
\end{figure}

In Fig.~\ref{fig:f15}, the dotted curves represent results of our computations with the combined
viscosity (\ref{eq:allvisc}) substituted into equation (\ref{eq:amt}). 
Note that a slightly increased amount of angular momentum that can be
transported from the core to the envelope on a longer hydrodynamical timescale has changed our results
in the right directions: first, after the wind constant $K_{\rm w}$ is properly recalibrated, the $P$-age relation
has shifted toward longer periods, and, second,  the degree of DR in the core has slightly been reduced.
To find out if these changes will further grow in the same directions when the efficiency of supplementary
angular momentum transport mechanisms increases, we have multiplied $\nu_{\rm GSF}$, the dominating
viscosity in the core, by a factor of 10. Results of this artificial viscosity enhancement are plotted in 
Fig.~\ref{fig:f15} with dashed curves. We see that both the $P$-age relation and the core rotation profile
have indeed continued to change in the right directions. This exercise shows that the revised prescription
cannot be rejected as easily as the original one. There remains a possibility that, when being
assisted by other transport processes, it may reproduce the observational data yet.

\begin{figure}
\epsfxsize=14cm
\epsffile [60 240 480 695] {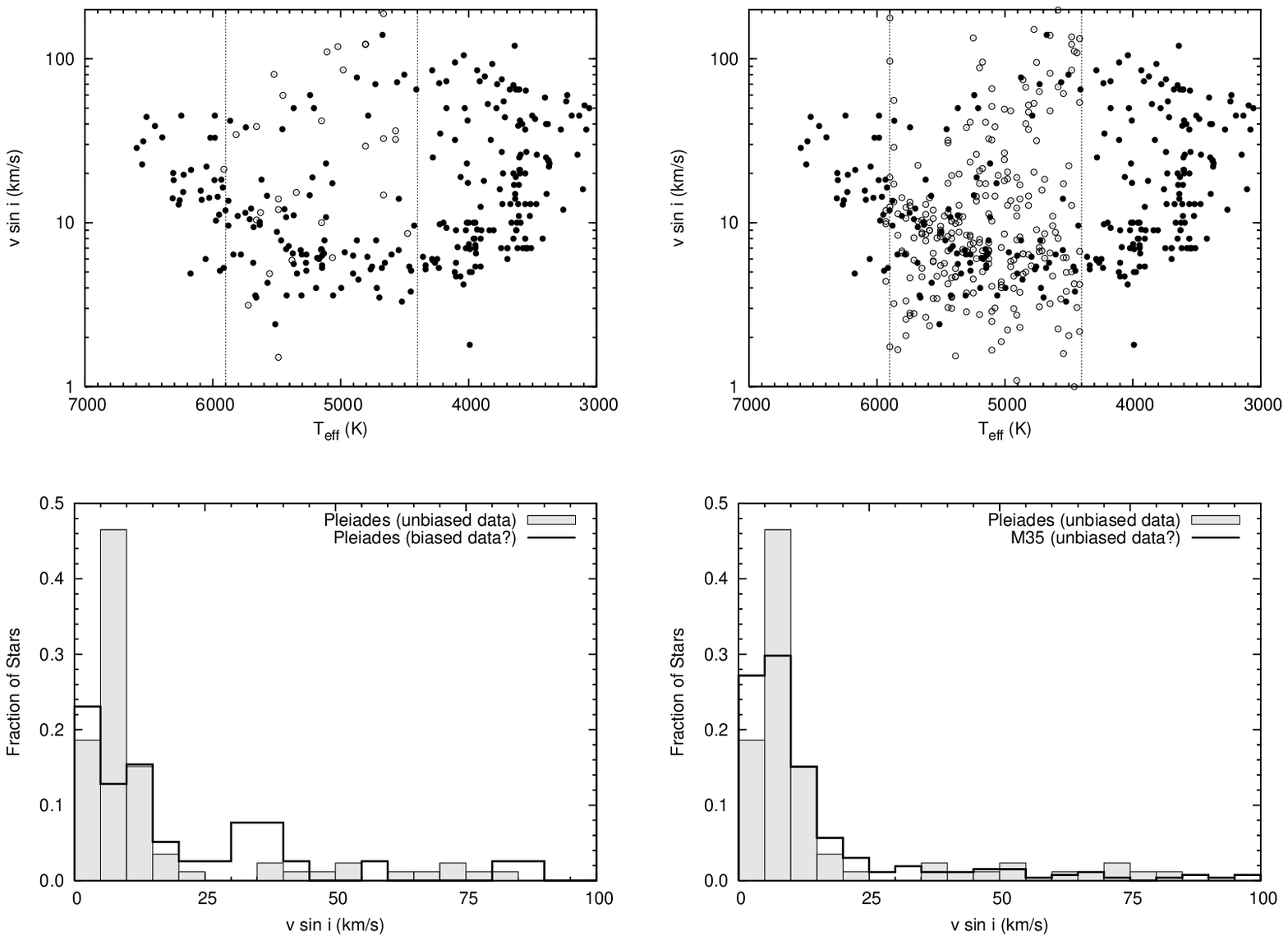}
\caption{Filled circles in upper panels are the Pleiades $v\sin i$ unbiased data
         from \cite{aps03}. Open circles in the upper left panel are the Pleiades data
         used in this paper that appear to be biased toward lower rotation periods. 
         Open circles in the upper right panel are
         the M\,35 data from \cite{mms09}. Periods have been transformed into $v\sin i$
         values using a randomly generated angle $0\leq i\leq \pi/2$ and $R = R_\odot$.
         The lower right panel shows that the unbiased data for the two clusters of similar age
         look alike ($P_{\rm KS} = 0.112$, the medians are 6.8 and 8.5, the first and third
         quartiles are 5.4 and 13 for the Pleiades, and 4.7 and 18 for M\,35).
         On the contrary, the biased Pleiades data in the lower left panel have very different median
         and third quartile values, 15 and 39, respectively. Vertical dotted lines in upper panels
         show the range of $0.7\la M/M_\odot\la 1.1$.
        } 
\label{fig:f16}
\end{figure}

Our main conclusion is critically based on a comparison of theoretical predictions and observations of
rotation periods for the slowest rotators in the intermediate-age open clusters.
Therefore, if the observational data were biased toward the longest periods that would undermine confidence
in our results. The real situation turns out to be opposite. Observations for some of our used open clusters
(for those with smaller data samples) are actually biased toward the shortest periods, as
is expected from the photometric period measurement procedure that needs longer observational times to accumulate
data sufficient for extracting longer periods. The following example illustrates this bias.
Filled circles in upper panels of Fig.~\ref{fig:f16} represent the Pleiades $v\sin i$ unbiased data
from \cite{aps03}. For comparison, open circles in the upper left panel show the Pleiades data
used in our work that do appear to be biased toward lower rotation periods. 
On the other hand, open circles in the upper right panel are
the M\,35 data from \cite{mms09}. Their original periods have been transformed into $v\sin i$
values using a randomly generated angle $0\leq i\leq \pi/2$ and $R = R_\odot$.
The lower right panel shows that the unbiased data for the two clusters of similar age
look alike ($P_{\rm KS} = 0.112$, the medians are 6.8 and 8.5, the first and third
quartiles are 5.4 and 13 for the Pleiades, and 4.7 and 18 for M\,35).
On the contrary, the biased Pleiades data in the lower left panel have very different median
and third quartile values, 15 and 39, respectively.

To sum up, our main conclusions can concisely be formulated as follows.
Whereas the period distributions for the fastest rotators among solar-type stars in open clusters are      
very well reproduced assuming their SB rotation, the period distributions for the slowest rotators are better 
described by stellar models with DR in their radiative cores. This conclusion is in agreement with previous
results reported by \cite{iea07}. Our new result is that the original prescription for the Tayler-Spruit dynamo
always enforces SB rotation in a solar-type star even if the star is a slow rotator. Therefore, this angular momentum
transport mechanism is unlikely to operate in solar-type early MS stars. The revised prescription
for the Tayler-Spruit dynamo cannot explain the observed period distributions either.
Besides, it leaves a large rapidly rotating radiative core in the present-day solar model. To be consistent
with observations, the revised prescription needs to be assisted by other angular momentum transport
mechanisms that must be able to penetrate into the deep core and operate on a longer timescale.

\acknowledgements
We are grateful to Luisa Rebull for providing us with the latest data on
rotation periods of open cluster low-mass stars.
We acknowledge support from the NASA grant NNG05 GG20G.


\clearpage

\clearpage


\clearpage


\clearpage


\clearpage


\clearpage


\clearpage
\begin{deluxetable}{cccccc}
\rotate
\tabletypesize{scriptsize}
\tablecolumns{6}
\tabletypesize{\footnotesize}
\tablecaption{Rotation Period Data}
\tablewidth{0pt}
\tablehead{
\colhead{} & \colhead{} & \multicolumn{3}{c}{Number of Stars in the Sample} & \colhead{} \\
\cline{3-5} \\
\colhead{Cluster} & \colhead{Age (Myr)} & \colhead{$M/M_\odot\leq 0.4$} & \colhead{$0.7\leq M/M_\odot < 0.9$} &
\colhead{$0.9\leq M/M_\odot\leq 1.1$} & \colhead{Source} }
   \startdata
   ONC & 2 &  32 & 3 & 53 & \cite{rws04} \\
   NGC\,2264 & 4 & 13 & 12 & 63 & \cite{rws04} \\
   Lupus & 4 & 0 & 2 & 12 & \cite{rws04} \\
   NGC\,2362 & 5 & 70 & 46 & 44 & \cite{iea08b} \\
   Tau-Aur & 6 & 0 & 2 & 23 & \cite{rws04} \\
   NGC\,2547 & 30 & 94 & 26 & 0 & \cite{iea08a} \\
   IC\,2391 \& IC\,2602 & 40 & 0 & 8 & 28 & \cite{rws04} \\
   $\alpha$\,Per & 70 & 0 & 8 & 28 & \cite{rws04} \\
   Pleiades & 115 & 0 & 20 & 19 & \cite{rws04} \\
   Pleiades ($v\sin i$) & 115 & 39 & 50 & 51 & \cite{aps03} \\
   M\,50 & 130 & 163 & 252 & 62 & \cite{iea09} \\
   NGC\,2516 & 180 & 183 & 0 & 0 & \cite{iea08b} \\
   M\,35 & 150 & 0 & 154 & 111 & \cite{mms09} \\
   M\,34 & 220 & 10 & 35 & 8 & \cite{iea06} \\
   M\,37 & 550 & 0 & 197 & 128 & \cite{hea09} \\
   Hyades & 600 & 0 & 8 & 17 & \cite{rea87,pea95} \\
   \enddata
\label{tab:tab1}
\end{deluxetable}



\begin{thebibliography}{}

\bibitem[Allain(1998)]{a98}
Allain, S.~1998, A\&A, 333, 629

\bibitem[Andronov, Pinsonneault, \& Sills(2003)]{aps03}
Andronov, N., Pinsonneault, M.~H., \& Sills, A.~2003, ApJ, 582, 358 

\bibitem[Alexander \& Ferguson(1994)]{af94}
Alexander, D.~R., \& Ferguson, J.~W.~1994, ApJ, 437, 879

\bibitem[Angulo et al.(1999)]{aea99}
Angulo, C., Arnold, M., Rayet, M., et al.~1999, NuPhA, 656, 3

\bibitem[Bouvier, Forestini, \& Allain(1997)]{bfa97}
Bouvier, J., Forestini, M., \& Allain, S.~1997, A\&A, 326, 1023

\bibitem[Bouwman et al.(2006)]{bea06}
Bouwman, J., Lawson, W.~A., Dominik, C., Feigelson, E.~D., Henning, T., Tielens, A.~G.~G.~M.,
\& Waters, L.~B.~F.~M.~2006, ApJ, 635, L57

\bibitem[Chanam\'{e}, Pinsonneault, \& Terndrup(2005)]{chpt05}
Chanam\'{e}, J., Pinsonneault, M., \& Terndrup, D.~M.~2005, ApJ, 631, 540 

\bibitem[Charbonneau \& MacGregor(1993)]{chmg93}
Charbonneau, P., \& MacGregor, K.~B.~1993, ApJ, 417, 762

\bibitem[Charbonnel \& Talon(2005)]{cht05}
Charbonnel, C., \& Talon, S.~2005, Science, 309, 2189

\bibitem[Couvidat et al.(2003)]{cea03}
Couvidat, S., Garc\'{\i}a, R.~A., Turck-Chi\`{e}ze, Corbard, T., Henney, C.~J.,
\& Jim\'{e}nez-Reyes, S.~2003, ApJ, 597, L77

\bibitem[Damjanov et al.(2007)]{dea07}
Damjanov, I., Jayawardhana, R., Scholz, A., Ahmic, M., Nguyen, D.~C., Brandeker, A., \&
van Kerkwijk, M.~H.~2007, ApJ, 670, 1337

\bibitem[Denissenkov \& VandenBerg(2003)]{dv03}
Denissenkov, P.~A., \& VandenBerg, D.~A.~2003, ApJ, 598, 1246

\bibitem[Denissenkov \& Pinsonneault(2007)]{dp07}
Denissenkov, P.~A., \& Pinsonneault, M.~2007, ApJ, 655, 1157

\bibitem[Denissenkov et al.(2008)]{dea08}
Denissenkov, P.~A., Pinsonneault, M., \& MacGregor, K.~B.~2008, ApJ, 684, 757

\bibitem[Eggenberger et al.(2005)]{eea05}
Eggenberger, P., Maeder, A., \& Meynet, G.~2005, A\&A, 440, L9

\bibitem[Garc\'{\i}a et al.(2007)]{gea07}
Garc\'{\i}a, R.~A., Turck-Chi\`{e}ze, S., Jim\'{e}nez-Reyes, S.~J.,
Ballot, J., Pall\'{e}, P.~L., Eff-Darwich, A., Mathur, S., \& Provost, J.~2007, Sience, 316, 1591

\bibitem[Grevesse \& Noels(1993)]{gn93}
Grevesse, N., \& Noels, A.~1993, in Origin and Evolution of the Elements,
ed. N. Prantzos, E. Vangioni-Flam, \& M. Casse (Cambridge: Cambridge Univ. Press), 15

\bibitem[Hartman et al.(2009)]{hea09}
Hartman, J.~D., Gaudi, B.~S., Pinsonneault, M.~H., Stanek, K.~Z., Holman, M.~J., 
McLeod, B.~A., Meibom, S., Barranco, J.~A., \& Kalirai, J.~S.~2009, ApJ, 691, 342

\bibitem[Herbst et al.(2007)]{hea07}
Herbst, W., Eisl\"{o}ffel, J., Mundt, R., \& Scholz, A., 2007,
in Protostars and Planets V, ied. B. Reipurth, D. Jewitt, and K. Keil 
(Tucson: University of Arizona Press), 297

\bibitem[Hillenbrand(2005)]{h05}
Hillenbrand, L.~A.,~2005, arXiv:astro-ph/0511083v1

\bibitem[Irwin et al.(2006)]{iea06}
Irwin, J., Aigrain, S., Hodgkin, S., Irwin, M., Bouvier, J., Clarke, C.,
Hebb, L., \& Moraux, E., 2006, MNRAS, 370, 954

\bibitem[Irwin et al.(2007)]{iea07}
Irwin, J., Hodgkin, S., Aigrain, S., Hebb, L., Bouvier, J., Clarke, C.,
Moraux, E., \& Bramich, D.~M., 2007, MNRAS, 377, 741

\bibitem[Irwin et al.(2008a)]{iea08a}
Irwin, J., Hodgkin, S., Aigrain, S., Bouvier, J., Hebb, L., \&
Moraux, E., 2008, MNRAS, 383, 1588

\bibitem[Irwin et al.(2008b)]{iea08b}
Irwin, J., Hodgkin, S., Aigrain, S., Bouvier, J., Hebb, L., Irwin, M., \&
Moraux, E., 2008, MNRAS, 384, 675 

\bibitem[Irwin et al.(2009)]{iea09}
Irwin, J., Aigrain, S., Bouvier, J., Hebb, L., Hodgkin, S., Irwin, M., \&
Moraux, E., 2009, MNRAS, 392, 1456

\bibitem[Itoh et al.(1996)]{iea96}
Itoh, N., Hayashi, H., Nishikawa, A., \& Kohyama, Y.~1996, ApJS, 102, 411

\bibitem[Jayawardhana et al.(2006)]{jea06}
Jayawardhana, R., Coffey, J., Scholz, A., Brandeker, A., \& van Kerkwij, M.~H.~2006, ApJ, 648, 1206

\bibitem[Keppens, MacGregor, \& Charbonneau(1995)]{kmc95}
Keppens, R., MacGregor, K.~B., \& Charbonneau, P.~1995, A\&A, 294, 469

\bibitem[Koenigl(1991)]{k91}
Koenigl, A.~1991, ApJ, 370, L39

\bibitem[Krishnamurthi et al.(1997)]{kea97}
Krishamurthi, A., Pinsonneault, M.~H., Barnes, S., \& Sofia, S.~1997, ApJ, 480, 303

\bibitem[Lyo \& Lawson(2005)]{ll05}
Lyo, A.-R., \& Lawson, W.~A.~2005, JKAS, 38, 241

\bibitem[MacGregor(1991)]{mcg91}
MacGregor, K.~B.~1991, in Angular Momentum Evolution of Young Stars,
ed. S. Catalano, \& J.~R. Stauffer, (Dordrecht: Kluwer Academic Publishers), 315

\bibitem[MacGregor \& Brenner(1991)]{mb91}
MacGregor, K.~B., \& Brenner, M..~1991, ApJ, 376, 204 

\bibitem[Maeder \& Zahn(1998)]{mz98}
Maeder, A., \& Zahn, J.-P..~1998, A\&A, 334, 1000

\bibitem[Mamajek(2009)]{m09}
Mamajek, E.~E.~2009, AIP Conf. Proceedings, vol. 1158, pp. 3-10, arXiv:0906.5011v1\,[astro-ph.EP]

\bibitem[Matt \& Pudritz(2005)]{mp05}
Matt, S., \& Pudritz, R.~E..~2005, ApJ, 632, L135

\bibitem[Meibom, Mathieu \& Stassun(2009)]{mms09}
Meibom, S., Mathieu, R.~D., \& Stassun, K.~G..~2009, ApJ, 695, 679

\bibitem[Mestel \& Weiss(1987)]{mw87}
Mestel, L., \& Weiss, N.~O..~1987, MNRAS, 226, 123 

\bibitem[Palacios et al.(2003)]{pea03}
Palacios, A., Talon, S., Charbonnel, C., \& Forestini, M.~2003, A\&A, 399, 603

\bibitem[Palla \& Stahler(1991)]{ps91}
Palla, F., \& Stahler, S.~W.~1991, ApJ, 375, 288

\bibitem[Pinsonneault, Kawaler, \& Demarque(1990)]{pkd90}
Pinsonneault, M.~H., Kawaler, S.~D., \& Demarque, P.~1990, ApJS, 74, 501

\bibitem[Prosser et al.(1995)]{pea95}
Prosser, C.~F., Shetrone, M.~D., Dasgupta, A., Backman, D.~E., Laaksonen, B.~D., Baker, S.~W.,
Marschall, L.~A., Whitney, B.~A., Kuijken, K., \& Stauffer, J.~R.~1995, PASP, 107, 211

\bibitem[Radick et al.(1987)]{rea87}
Radick, R.~R., Thompson, T.~D., Lockwood, G.~W., Duncan, D.~K., \& Baggett, W.~E.~1987, ApJ, 321, 459

\bibitem[Rebull, Wolff, \& Strom(2004)]{rws04}
Rebull, L.~M., Wolff, S.~C., \& Strom, S.~E.~2004, AJ, 127, 1029

\bibitem[Rogers \& Iglesias(1992)]{ri92}
Rogers, F.~J., \& Iglesias, C.~A.~1992, ApJ, 401, 361

\bibitem[R\"{u}diger, Gellert, \& Schultz(2009)]{rgs09}
R\"{u}diger, G., Gellert, M., \& Schultz, M.~2009, MNRAS, 399, 996 

\bibitem[Scholz \& Eisl\"{o}ffel(2007)]{se07}
Scholz, A., \& Eisl\"{o}ffel, J.~2007, MNRAS, 381, 1638

\bibitem[Shu et al.(1994)]{shea94}
Shu, F., Najita, J., Ostriker, E., Wilkin, F., Ruden, S., \& Lizano, S.          
1994, ApJ, 429, 781

\bibitem[Sicilia-Aguilar et al.(2009)]{saea09}
Sicilia-Aguilar, A., Bouwman, J., Juh\'{a}sz, A., Henning, T., Roccatagliata, V., Lawson, W.~A.,
Acke, B., Feigelson, E.~D., Tielens, A.~G.~G.~M., Decin, L., \& Meeus, G.~2009, ApJ, 701, 1188
1994, ApJ, 429, 781

\bibitem[Siess \& Livio(1997)]{sl97}
Siess, L., \& Livio, M.~1997, ApJ, 490, 785

\bibitem[Sills, Pinsonneault, \& Terndrup(2000)]{spt00}
Sills, A., Pinsonneault, M.~H., \& Terndrup, D.~M.~2000, ApJ, 534, 335

\bibitem[Spruit(1999)]{s99}
Spruit, H.~C.~1999, A\&A, 349, 189

\bibitem[Spruit(2002)]{s02}
Spruit, H.~C.~2002, A\&A, 381, 923

\bibitem[Tayler(1973)]{t73}
Tayler, R.~J.~1973, MNRAS, 161, 365

\bibitem[Tinker, Pinsonneault, \& Terndrup(2002)]{tpt02}
Tinker, J., Pinsonneault, M., \& Terndrup, D.~M.~2002, ApJ, 564, 877

\bibitem[Tomczyk, Schou, \& Thompson(1995)]{tst95}
Tomczyk, S., Schou, J., \& Thompson, M.~J.~1995, ApJ, 448, L57

\bibitem[Zahn(1992)]{z92}
Zahn, J.-P.~1992, A\&A, 256, 115

\bibitem[Zahn, Brun, \& Mathis(2007)]{zbm07}
Zahn, J.-P., Brun, A.~S., \& Mathis, S.~2007, A\&A, 474, 145

\end{thebibliography}
\end{document}